%% file: ckm16-wg3.tex
\documentclass{PoS}
\usepackage{url}

\title{Summary of the CKM 2016 working group on rare decays}

\ShortTitle{Summary of the CKM 2016 working group on rare decays}

\author{Akimasa Ishikawa\\
        Department of Physics, Tohoku University, Sendai, Japan\\
        E-mail: \email{akimasa@epx.phys.tohoku.ac.jp}}

\author{Enrico Lunghi\\
        Physics Department, Indiana University, Bloomington IN, USA\\
        E-mail: \email{elunghi@indiana.edu}}

\author{Matthew Moulson\\
        INFN Laboratori Nazionali di Frascati, Italy\\
        E-mail: \email{Matthew.Moulson@lnf.infn.it}}

\author{Justine Serrano\\
        Aix Marseille Univ, CNRS/IN2P3, CPPM, Marseille, France\\
        E-mail: \email{serrano@cppm.in2p3.fr}}

\abstract{\input{abstract}}

\FullConference{9th International Workshop on the CKM Unitarity Triangle\\
		28  November - 3 December 2016\\
		Tata Institute for Fundamental Research (TIFR), Mumbai, India}

\makeatletter
\g@addto@macro\bfseries{\boldmath}
\makeatother

\begin{document}

\input{introduction}

\input{rareb}

\input{rarec}
\input{rarek}
\input{chargeLFV}
\input{summary}

\end{document}

%% file: abstract.tex
Rare $B$, $D$, and $K$ decays offer unique opportunities to probe for
evidence of new particles from physics beyond the Standard Model
at mass scales extending from the electroweak scale to well above
those directly accessible at the LHC. We review a selection of
theoretical and experimental results on rare $B$, $D$, and $K$ decays
illustrating the progress made during the past two years.

%% file: introduction.tex
\section{Introduction}
The presence of a single Higgs doublet in the Standard Model of particle physics (SM) implies that all flavour-changing transitions are determined by the three angles and the one phase that parameterize the Cabibbo-Kobayashi-Maskawa matrix (CKM)~\cite{Cabibbo:1963yz, Kobayashi:1973fv}. Rare $B$, $D$, and $K$ decays proceed at loop level via vertices suppressed by small, off-diagonal CKM entries, thus offering exquisite tests of the Higgs mechanism of electroweak symmetry breaking. 

Deviations from SM predictions are usually parameterized in terms of non-standard contributions to various non-renormalizable operators whose intrinsic scale cannot be too far removed from the electroweak scale in order to produce observable effects. While this usually means that non-standard contributions to rare decays are associated with new particles that are within the reach of the LHC, there are some observables (especially involving kaons) that are sensitive to energy scales that are far beyond current collider capabilities.

During the past two years there has been enormous experimental progress, with ground breaking analyses from LHCb, ATLAS, CMS, BaBar, and Belle. The general picture that emerges is one of overall agreement with SM expectations. However, there are several observables for which measurements are in some tension with the SM. Confirmation of these discrepancies would point to the existence of new massive particles that are clearly within the reach of ATLAS and CMS and which would very likely introduce novel non-CKM like flavour changing interactions. The most notable anomalies appear in $B\to K^{(*)} \ell\ell$ and $B\to D^{(*)}\ell\nu$ decays. 

The upcoming high luminosity $B$-factory Belle II is about to come online, promising important cross checks of these effects~\cite{belleII}.  Moreover, the NA62~\cite{NA62:2017rwk} and KOTO~\cite{Komatsubara:2012pn,Ahn:2016kja} experiments are also taking data and will offer precise measurements of the extremely rare kaon decays $K^+ \to \pi^+\nu\bar \nu$ and $K_L\to\pi^0\nu\bar\nu$, whose branching ratios can be calculated with incredible theoretical accuracy.

In the following, we review some of the recent theoretical and experimental progress that has been made during the last two years in rare $B$, $D$, and $K$ decays.

%% file: rareb.tex
\section{Rare \boldmath{$b$}-hadron decays}
Rare FCNC decays of $b$-hadrons can be described in a model-independent approach using the effective Hamiltonian
\begin{equation}
\mathcal{H}_{\rm eff}= -\frac{4G_F}{\sqrt{2}}V_{tb}V_{tq}^*\sum\limits_{i} (C_i\mathcal{O}_i+C^\prime_i\mathcal{O}^\prime_i)+h.c.,
\end{equation}
where $q=d,s$ for processes based on the quark level $b\to d,s$ transitions. The heavy degrees of freedom have been integrated out in the short distance Wilson coefficients $C_i$, and the operators $\mathcal{O}_i$ encode the long-distance effects. In the SM, the main operators are the electromagnetic operator  $\mathcal{O}_7$  and the semileptonic operators  $\mathcal{O}_9$ and $\mathcal{O}_{10}$. New-physics (NP) contributions could affect the value of the Wilson coefficients $C_{7,9,10}$ or involve other operators such as  $\mathcal{O}^\prime_{7,9,10}$ or $\mathcal{O}^{(\prime)}_{S,P}$. The different $\mathcal{O}_i$ contribute differently to leptonic, semileptonic, and radiative decays. The next three sections present the experimental results regarding these channels, while Section~\ref{sec:global} presents a global analysis of these measurements in the effective Hamiltonian framework.
\input{leptonic}

\input{semileptonic}

\input{radiative}
\input{global_analysis}

%% file: leptonic.tex
\subsection{Leptonic decays}

The $B^0_d \to \mu^+ \mu^- $ and $B^0_s \to \mu^+ \mu^- $ channels  are particularly sensitive to NP contributions in the scalar/pseudoscalar sector 
and have been searched for for more than 25 years. The results of a combined analysis of the CMS and LHCb Run 1 data were presented at CKM 2014, revealing the first observation of the $B^0_s \to \mu^+ \mu^-$ decays. The measured  branching ratios are compatible with SM expectations, ${\rm BR}(B^0_s \to \mu^+ \mu^- )= (3.65\pm0.23)\times 10^{-9}$ and ${\rm BR}(B^0_d \to \mu^+ \mu^- )=(1.06\pm0.09)\times 10^{-10} $ \cite{Bobeth:2013uxa},  at the 1.2 $\sigma$ level for the $B^0_s$ and 2.2$\sigma$  for the $B^0_d$ decays. 
%and the first evidence for the 
%$B^0 \to \mu^+ \mu^- $ decay, with branching fractions compatible with the SM expectations at 1.2 and 2.2$\sigma$ level, respectively.
The ATLAS collaboration recently presented the results of its analysis of Run~1 data,  ${\rm BR}(B^0_s \to \mu^+ \mu^-)= (0.9^{+1.1}_{-0.8})\times 10^{-9}$ and 
${\rm BR}(B^0 \to \mu^+ \mu^-) < 4.2\times 10^{-10}$ at 95\% CL \cite{Aaboud:2016ire}, which are in agreement with the CMS+LHCb combination.
The analysis of the data taken by the LHC experiments during Run 2 will improve these measurements and start to provide additional observables such as the ratio of the $B^0_d$ and $B^0_s$ modes, which is sensitive to Minimal Flavour Violation scenarios, or the $B^0_s \to \mu^+ \mu^- $ effective lifetime, which has different sensitivity to NP models with scalar and non-scalar contributions.

Searches for leptonic $B$ decays into $\tau$ leptons are interesting in view of the recent hints of lepton flavour non-universality obtained by several experiments. 
Their branching ratios are two orders of magnitude higher than those for decays into muons because of the less stringent helicity suppression and the higher lepton mass: 
 ${\rm BR}(B^0_s \to \tau^+ \tau^- )= (7.73\pm0.49)\times 10^{-7}$ and ${\rm BR}(B^0_d \to \tau^+ \tau^- )=(2.22\pm0.19)\times 10^{-8} $ \cite{Bobeth:2013uxa}. However, experimental searches for these decays are complicated by the  presence of at least two neutrinos in the final state from the $\tau$ decays.
The LHCb Collaboration has presented preliminary results corresponding to the first limit on BR($B^0_s \to \tau^+ \tau^-$), at $3.0\times 10^{-3}$ (95\% CL), and the best limit on ${\rm BR}(B^0_d \to \tau^+ \tau^-)$, at $1.3\times 10^{-3}$ (95\% CL) \cite{DeBruyn:2016tiq}, obtained using the hadronic tau decay $\tau^+\to\pi^-\pi^+\pi^-\nu_{\tau}$.

%% file: semileptonic.tex
\subsection{Semileptonic decays}

Rare  semileptonic decays provide two classes of observables allowing tests of the SM. The observables in the first class provide tests of lepton flavour universality and are theoretically very clean. An example is the measurement of the ratio $R_K=\Gamma(B^+\to K^+ \mu^+\mu^-) / \Gamma(B^+\to K^+ e^+ e^-)$, which was found to be $R_K=0.745^{+0.090}_{-0.074}\pm0.036$ by  LHCb in the range $1< q^2 <6~ \mathrm{GeV}^2/c^4$ \cite{Aaij:2014ora}, 2.6$\sigma$ lower than the SM prediction, $1.00\pm0.01$ \cite{Bordone:2016gaq} ($q^2$ is the squared invariant mass of the dilepton). A new BaBar analysis performed in the same $q^2$ region confirmed the deficit in the muonic mode, measuring  $R_K=0.64^{+0.39}_{-0.30}\pm0.06$. 

The observables in the second class include differential branching ratios and  angular distributions. Their theoretical predictions are affected by hadronic uncertainties arising from the form factors, which are computed using lattice QCD or light cone sum-rule techniques, depending on the $q^2$ region.
An interesting picture emerges from the differential branching ratio measurements for
$b \to s\mu^+ \mu^-$ decays from the LHCb Collaboration, in which, below the charmonium resonances,  the experimental values tend to be lower than the SM predictions \cite{Aaij:2014pli,Aaij:2015xza,Aaij:2015esa,Aaij:2016flj}, in agreement with the deficit in the muonic mode seen in $R_K$.  The largest effect is seen in the $B_s \to \phi\mu^+ \mu^-$ 
channel, where the discrepancy is at the level of 3$\sigma$. 

%%%%%%%%%%%%%%%%%%%%
\begin{figure}[!h]
\center
\includegraphics[width=0.49\textwidth]{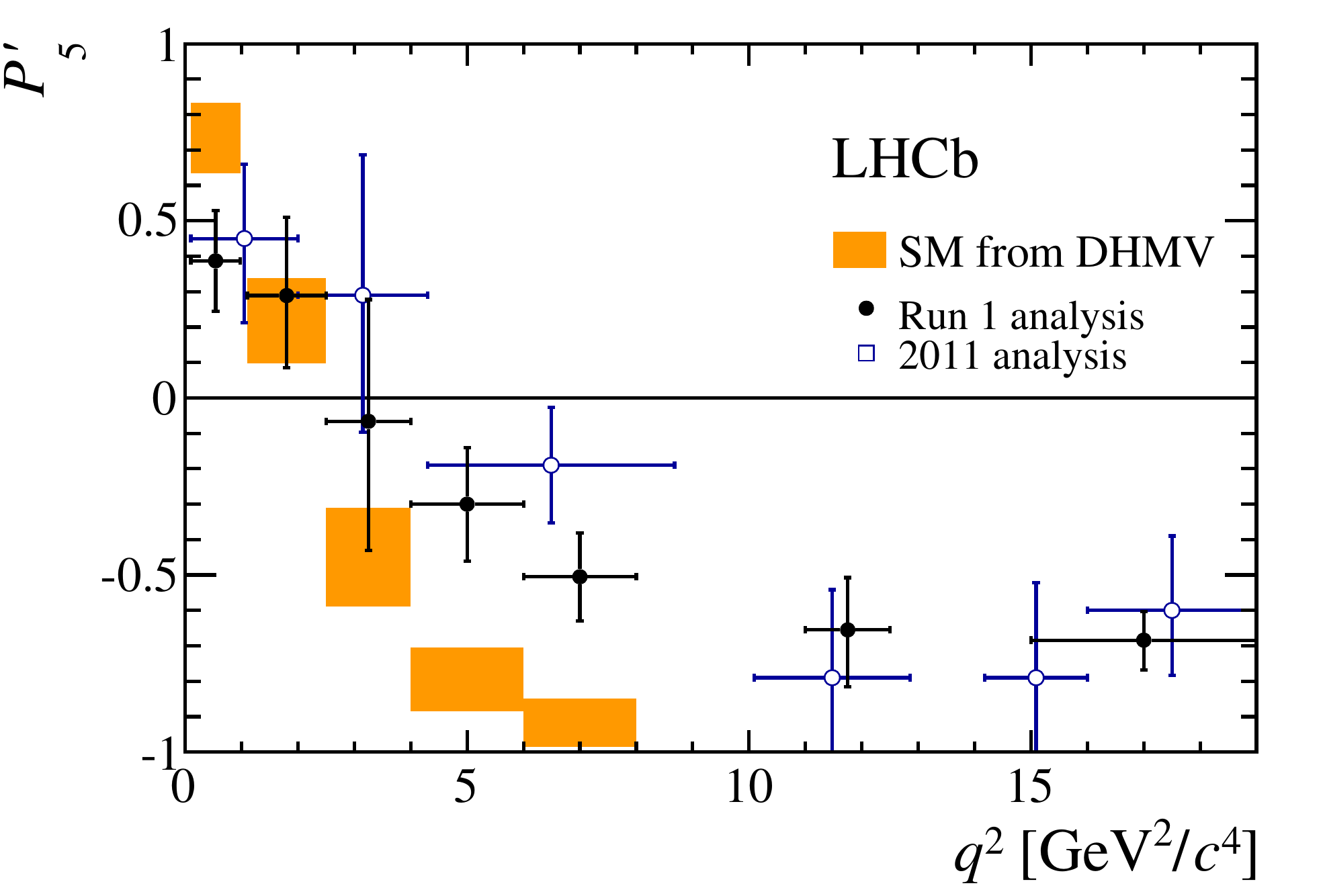}
\includegraphics[width=0.49\textwidth]{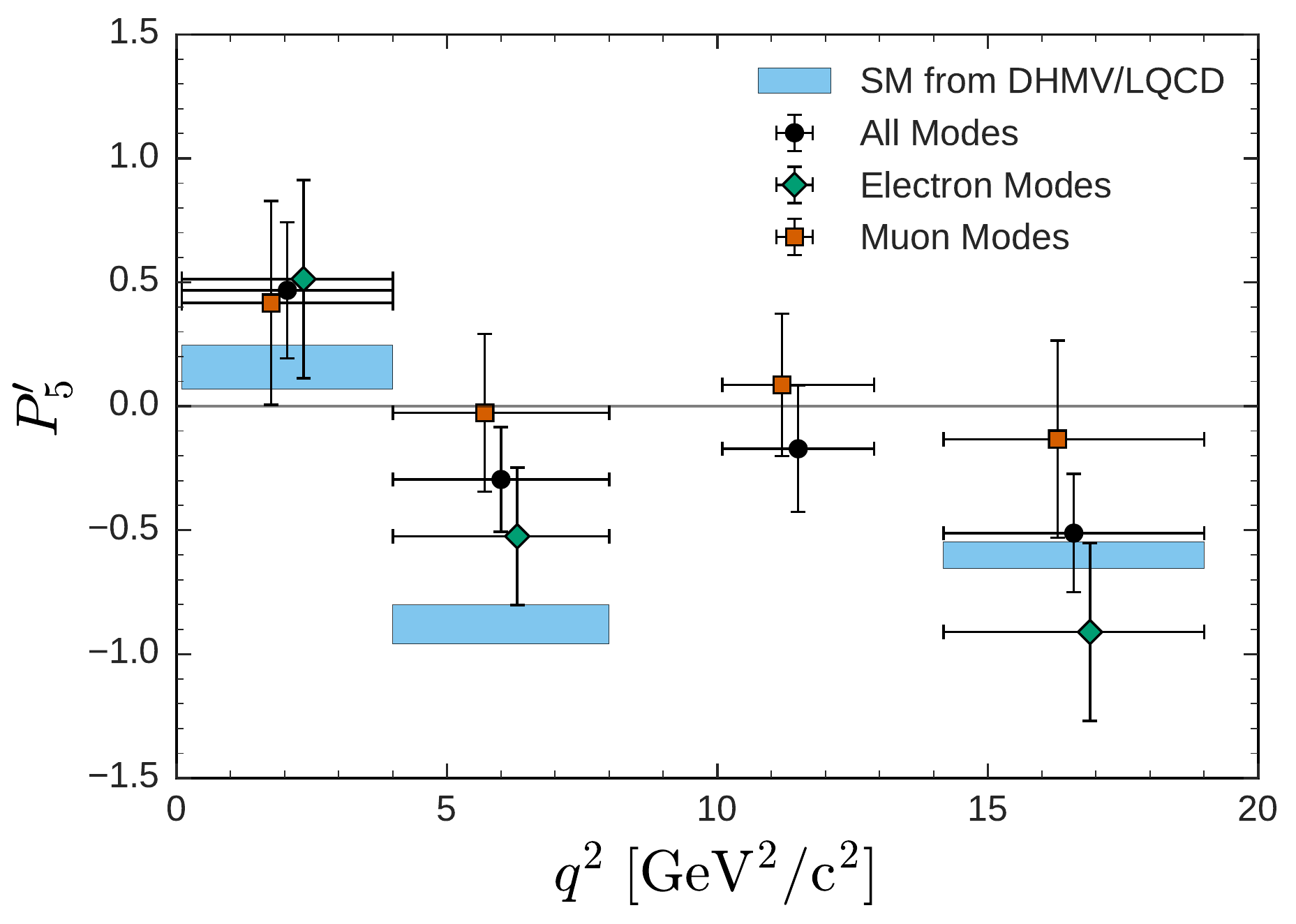}
\caption{ The $P'_5$ angular observables in bins of $q^2$. The shaded boxes show the SM prediction taken from \cite{Descotes-Genon:2014uoa}. Left: LHCb results from $B^0 \to K^{*0}\mu^+ \mu^-$ data. Right: Belle results from $B^{(+)} \to K^{*(+)}\ell^+ \ell^-$ data for electron modes, muon modes, and the combination. } 

\label{Fig:LHCb_P5}
\end{figure}
%%%%%%%%%%%%%%%%%%%%

The angular observables for  $B \to K^*\ell^+ \ell^-$ decays, where $\ell=e,\mu$,  have been studied by several experiments \cite{Lees:2015ymt,Khachatryan:2015isa,Aaij:2015oid,Wehle:2016yoi}. 
The differential decay rate can be described by 
\begin{equation}
\frac{1}{d\Gamma/dq^2}\cdot\frac{d^4\Gamma[\bar B^0+B^0]}{d \cos\theta_\ell\, d\cos\theta_K\, d\phi\, dq^2} = \frac{9}{32\pi}\sum_i S_i(q^2)f_i(\cos\theta_\ell,\cos\theta_K,\phi),
\end{equation}
where the observables $S_i$ are functions of the Wilson coefficients.
Other observables with reduced sensitivity to the hadronic uncertainties can be formed from the  $S_i$, in particular 
$P'_5=\frac{S_5}{\sqrt{F_L(1-F_L)}}$, where $F_L$ is the fraction of longitudinally polarized $K^*$s.

Using the Run 1 data set of 3~$\mathrm{fb}^{-1}$, the LHCb Collaboration has performed a full angular analysis of the $B^0$ decay, confirming the tension
previously seen with 1~$\mathrm{fb}^{-1}$ between the measurements of the $P'_5$ observables and SM predictions at low $q^²$, as shown in Fig.~\ref{Fig:LHCb_P5}, left \cite{Aaij:2015oid}. This tension is now at the 3.4$\sigma$ level.
The Belle collaboration has used a folding technique to access the same observable, combining the $B^0$ and $B^+$ modes, and reported  a tension of 2.6$\sigma$ in the bin $4 < q^2 < 8~\mathrm{GeV}^2/c^4$   \cite{Wehle:2016yoi}.
They also measured for the first time the $Q_i= P^\mu_i-P^e_i$ variables introduced in \cite{Capdevila:2016ivx}, which also provide a test of lepton flavour universality.

The anomalies in $R_K$ and $P_5'$, together
with the pattern of neutrino mixings, can be simultaneously
explained by new physics in the operators $\mathcal{O}_9^\ell$
and $\mathcal{O}_9^{\prime\ell}$. The new physics can be introduced via the
$Z'$ of a $U(1)_X$ symmetry~\cite{Bhatia:2017tgo}.
A model can be constructed that satisfies the constraints from
$B$ and $K$ mixing and rare $B$ decays, as well as from direct searches
for $pp\to Z'\to\mu\mu$ at colliders. 
Direct detection of a 4~TeV $Z'$ in the $\mu\mu$ channel at the LHC would
require several hundred ${\rm fb}^{-1}$.

%K*nunu

\begin{figure}[!h]
\center
\includegraphics[width=0.6\textwidth]{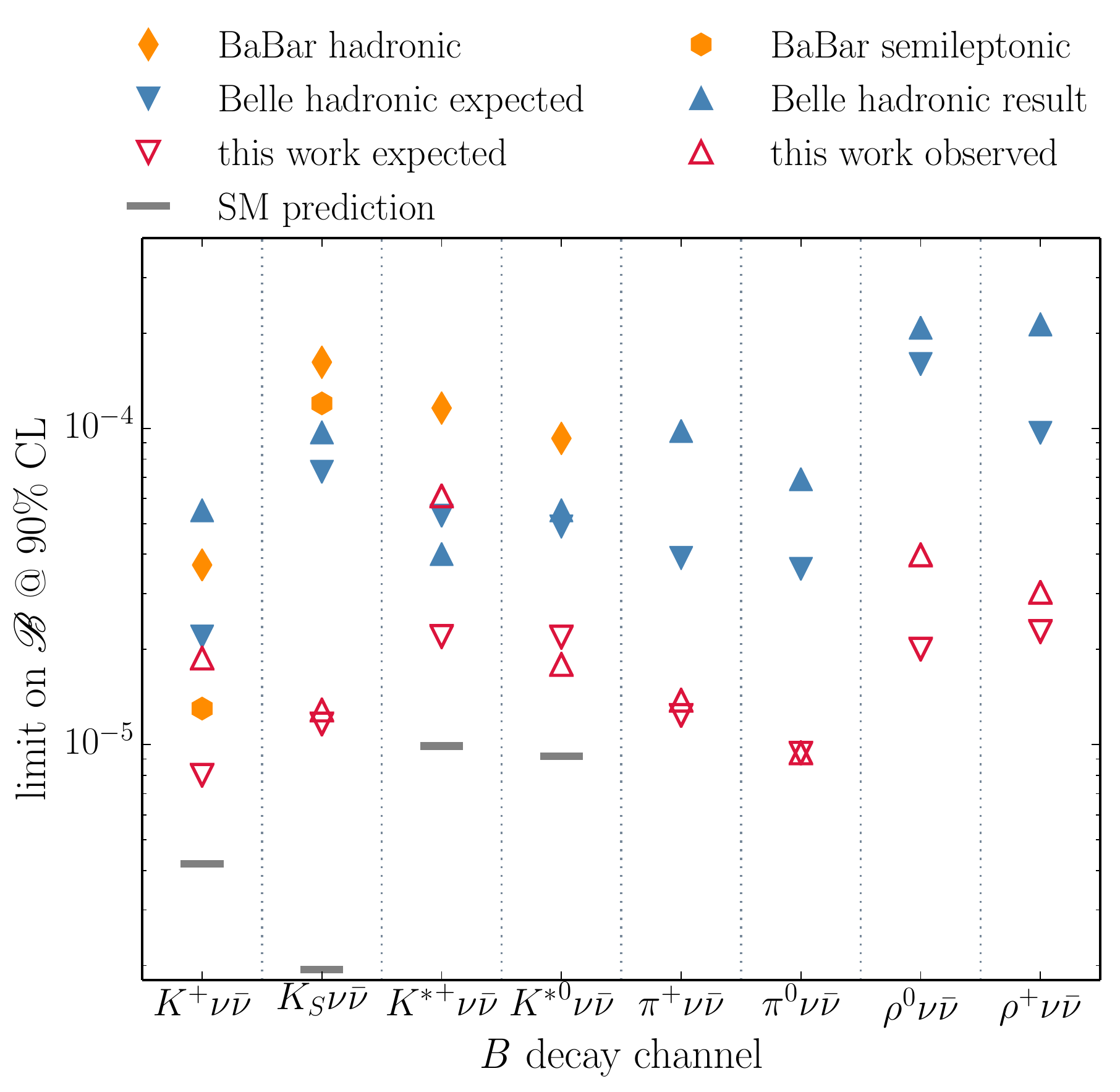}
\caption{  Expected and observed limits obtained in \cite{Grygier:2017tzo}  in comparison to previous results for the BaBar measurement with semileptonic \cite{delAmoSanchez:2010bk} and hadronic tag \cite{Lees:2013kla}, as well as the Belle measurement utilizing hadronic tagging \cite{Lutz:2013ftz}. Theory predictions are from \cite{Buras:2014fpa}. } 
\label{Fig:hnunu}
\end{figure}

The  $b \to (s,d) \nu \bar\nu$ transitions are theoretically cleaner than the modes with charged leptons, as only the $Z$ boson can intervene in the penguin diagram. According to the SM, the predicted branching ratios span a range from 
$2.4\times 10^{-7}$ for the $\pi^+$ mode \cite{Hambrock:2015wka} to $9.2\times 10^{-6}$ for the $K^{*+}$ mode \cite{Buras:2014fpa}.
Searches for these decay channels have been  performed by the Belle and BaBar experiments using full-event reconstruction thanks to hadronic  \cite{Lutz:2013ftz,Lees:2013kla} or semileptonic tagging modes \cite{delAmoSanchez:2010bk}.
At this conference, the Belle Collaboration presented new results based on an improved semileptonic tagging method leading
to the most stringent limits on $B^0 \to K_s^0 \nu \bar\nu$, $B^0 \to K^{*0} \nu \bar\nu$ , $B^+ \to \pi^+ \nu \bar\nu$, $B^0 \to \pi^0 \nu \bar\nu$ , $B^+ \to \rho^+ \nu \bar\nu$ and $B^0 \to \rho^0 \nu \bar\nu$ \cite{Grygier:2017tzo}, as seen in Fig. \ref{Fig:hnunu}. 
%%%%%%%%%%%%%%%%%%%%
%%%%%%%%%%%%%%%%%%%%
It should be noted that numerous other new results on $b\to (s,d)\ell\ell$ transitions have appeared since the last CKM workshop in 2014, such as the first observation of a $b\to d$ baryonic transition \cite{Aaij:2017ewm}, angular analysis of  $B_s \to \phi\mu^+ \mu^-$ \cite{Aaij:2015esa} and  $\Lambda_b \to \Lambda\mu^+ \mu^-$ decays \cite{Detmold:2016pkz}, the first search for  $B^+ \to K^+\tau^+ \tau^-$ \cite{TheBaBar:2016xwe}, and many more that cannot be detailed in this report.

%% file: radiative.tex
\subsection{Radiative decays}
The inclusive radiative $\bar B\to X_{s,d} \gamma$ decays have been the subject of intense studies over the last few decades. On the theoretical side the dominant contributions to these decays are described by a local operator product expansion (OPE) known at NNLO~\cite{Misiak:2015xwa} (with the exception of the exact $m_c$ dependence of the ${\cal O}_{1,2}$-${\cal O}_7$ interference~\cite{Czakon:2015exa, Misiak:2017woa}). Subdominant corrections appear at the power-suppressed level ($\sim\Lambda_{\rm QCD}/m_b$) and can be divided into local and non-local. The former are known through $O(\alpha_s \Lambda_{\rm QCD}^2/m_b^2)$~\cite{Ewerth:2009yr} and $O(\Lambda_{\rm QCD}^5/m_b^5)$~\cite{Gambino:2016jkc}, albeit with poor knowledge of the higher-power matrix elements. The latter lead to resolved photon contributions in which the photon emission is a long-distance rescattering effect (e.g. $b \to sc \bar c \to s g \gamma$)~\cite{Benzke:2010js}. Calculations of these effects are under very poor theoretical control and are essentially used to set an upper limit of about 5\% on the their possible size; this is the last single source of uncertainty on the theoretical prediction for these branching ratios. The current theoretical predictions are~\cite{Misiak:2015xwa}: ${\rm BR} (\bar B \to X_s \gamma)_{E_\gamma>1.6\; {\rm GeV}}^{\rm SM} =  (3.36 \pm 0.23)\times 10^{-4}$ and ${\rm BR} (\bar B \to X_d \gamma)_{E_\gamma>1.6\; {\rm GeV}}^{\rm SM} = \left(1.73^{+0.12}_{-0.22}\right)\times 10^{-5}$. Recently Belle presented an updated measurement using their full $711\; {\rm fb}^{-1}$ data set~\cite{Belle:2016ufb}: ${\rm BR} (\bar B \to X_s \gamma)_{E_\gamma>1.6\; {\rm GeV}}^{\rm exp} =  (3.12 \pm 0.10_{\rm stat} \pm 0.19_{\rm syst}\pm 0.08_{\rm model} )\times 10^{-4}$. The accuracy of this single measurement is identical to that of the previous world average. The uncertainty on this measurement is already dominated by systematics; however, the systematic uncertainty can be reduced by further studies with a larger data set. A total uncertainty of 3.2\% is reachable with 50~ab${}^{-1}$ at Belle~II.
% AI: 3.2% is the latest estimates

The situation is radically different for exclusive modes, where only the magnetic-moment operator contribution is described in terms of a local OPE, requiring the tensor $B\to (K^*, \rho,\ldots)$ and $B_s \to (\phi, \bar K^*, \ldots)$ form factors. Contributions of other operators can be calculated within the QCD factorization approach up to non-local power corrections. The latter introduce very sizable uncertainties that can be somewhat reduced by considering ratios and asymmetries like ${\rm BR}(B\to K^*\gamma)/{\rm BR}(B_s\to \phi\gamma)$ and the $B\to (K^*,\rho)\gamma$ isospin asymmetries, which are predicted with uncertainties of 23\% and 54\%, respectively~\cite{Lyon:2013gba}. Other observables, like the time-dependent $CP$ asymmetry measured in $B^0,\,\bar B^0\to f_{CP}\gamma$ are expected to be vanishingly small in the SM due to the chiral nature of weak interactions and offer sensitive tests of non-standard right-handed currents. In this context, a result for the time-dependent $CP$ asymmetry in the $K^0_S\rho\gamma$ final state has been recently obtained by BaBar~\cite{Sanchez:2015pxu}, resulting in $S_{K^0_S\rho\gamma} = -0.18 \pm 0.32^{+0.06}_{-0.05}$, which is compatible with the SM expectation ($S \sim 0.02$). The time-dependent decay rate of untagged $B_s \to \phi \gamma$ is also sensitive to the photon polarization via the coefficient of the $\sinh$ term, $A^{\Delta}$~\cite{Muheim:2008vu}. LHCb first measured this observable, obtaining $A^{\Delta} = -0.98^{+0.46}_{-0.52}{}^{+0.23}_{-0.20}$~\cite{Aaij:2016ofv}, which is consistent with the SM prediction from~\cite{Muheim:2008vu}, $A^{\Delta}=0.0047^{+0.029}_{-0.025}$.

%% file: global_analysis.tex
\subsection{Global analysis of $b\to s$ decays}
\label{sec:global}
Global fits to $b\to s$ anomalies involve the combination of several exclusive and inclusive $b\to s \gamma$ and $b\to s\ell\ell$ transitions: $B\to (K^*, X_s)\gamma$ (sensitive to $C_7^{(\prime)}$), $B_s\to \ell\ell$ (sensitive to $C_{10}^{(\prime)}$), $B\to (K,K^*,X_s)\ell\ell$, and $B_s\to \phi \ell\ell$ (sensitive to $C_{7,9,10}^{(\prime)}$). These studies revolve around the experimental tensions in exclusive $b\to s \ell\ell$ decays and depend critically on theoretical systematic uncertainties.

The decays $B\to (K,K^*)\ell\ell$ and $B_s\to \phi\ell\ell$ are described using Soft-Collinear Effective Theory (SCET)~\cite{Bauer:2000yr, Beneke:2001at} at low $q^2$, where the final-state hadron has large energy, and by a local OPE~\cite{Grinstein:2004vb, Bobeth:2010wg} at high $q^2$, where the final-state hadron is almost at rest. 

At low $q^2$, the leading contributions to the amplitudes are expressed in terms of heavy-to-light $B_q$ form factors and meson light-cone distribution amplitudes. Unfortunately, there is no widespread agreement on the actual size of the sub-leading corrections, which are expected to scale as $\Lambda_{\rm QCD}/m_b$.
For the calculation of branching ratios, it is always possible to use full QCD form factors, thus confining power corrections to the matrix elements of the operators ${\cal O}_{1,2}$ whose contribution to the total amplitude is subdominant (see for instance the $B\to K\ell\ell$ analysis presented in Ref.~\cite{Du:2015tda}). The main problem resides in the calculation of asymmetries and ratios (e.g. $P_5^\prime$) for which the leading dependence on the form factors cancels as long as the form factors themselves are calculated within SCET; this introduces additional power corrections, about the size of which there is currently no definite agreement~\cite{Descotes-Genon:2015uva, Jager:2012uw, Jager:2014rwa, Ciuchini:2015qxb}. At high $q^2$, the power corrections are local and thus under better control. Unfortunately, resonant charmonium contributions ($B\to  K^{(*)} \psi_{cc}\to  K^{(*)} \ell\ell$) introduce potential violations of quark-hadron duality that are difficult to estimate~\cite{Beylich:2011aq, Lyon:2014hpa}. 

The form factors are the most important non-perturbative inputs to these calculations and are accessible  using lattice QCD and Light-Cone QCD Sum Rules (LCSR). Lattice QCD offers a first-principle calculation in which all sources of uncertainty can be systematically taken into account, but, for technical reasons, allows access to the form factors only at relatively large momentum transfer ($q^2$). The LCSR approach, on the other hand, requires the final-state mesons to have large energy, implying small $q^2$. For these reasons it is common to see analyses using lattice QCD at low recoil and LCSR at high recoil. Recently, very high quality calculations of the three form factors $f_{+,0,T}(q^2)$ for the $B\to \pi$~\cite{Flynn:2015mha, Lattice:2015tia, Bailey:2015nbd}, $B_s\to K$~\cite{Bouchard:2014ypa, Flynn:2015mha}, and $B\to K$~\cite{Bouchard:2013pna, Bouchard:2013mia, Bailey:2015dka} channels have been performed. Decays into vector mesons are considerably more complex because there are seven independent form factors for each channel and the vector mesons undergo strong decays. A first complete study of $B\to K^*$ and $B_s \to (\phi,K^*)$ was presented in Ref.~\cite{Horgan:2013hoa}. Bottom baryon form factors have been also investigated in the $\Lambda_b \to (p,\Lambda_c)$~\cite{Detmold:2015aaa} and $\Lambda_b \to \Lambda$~\cite{Detmold:2016pkz} channels. Calculations of $B\to (K^*,\rho,\omega)$ and $B_s\to (\phi, K^*)$ in the LCSR approach are presented in Ref.~\cite{Straub:2015ica}.

\begin{figure}[t]
\center
\includegraphics[width=0.49\textwidth]{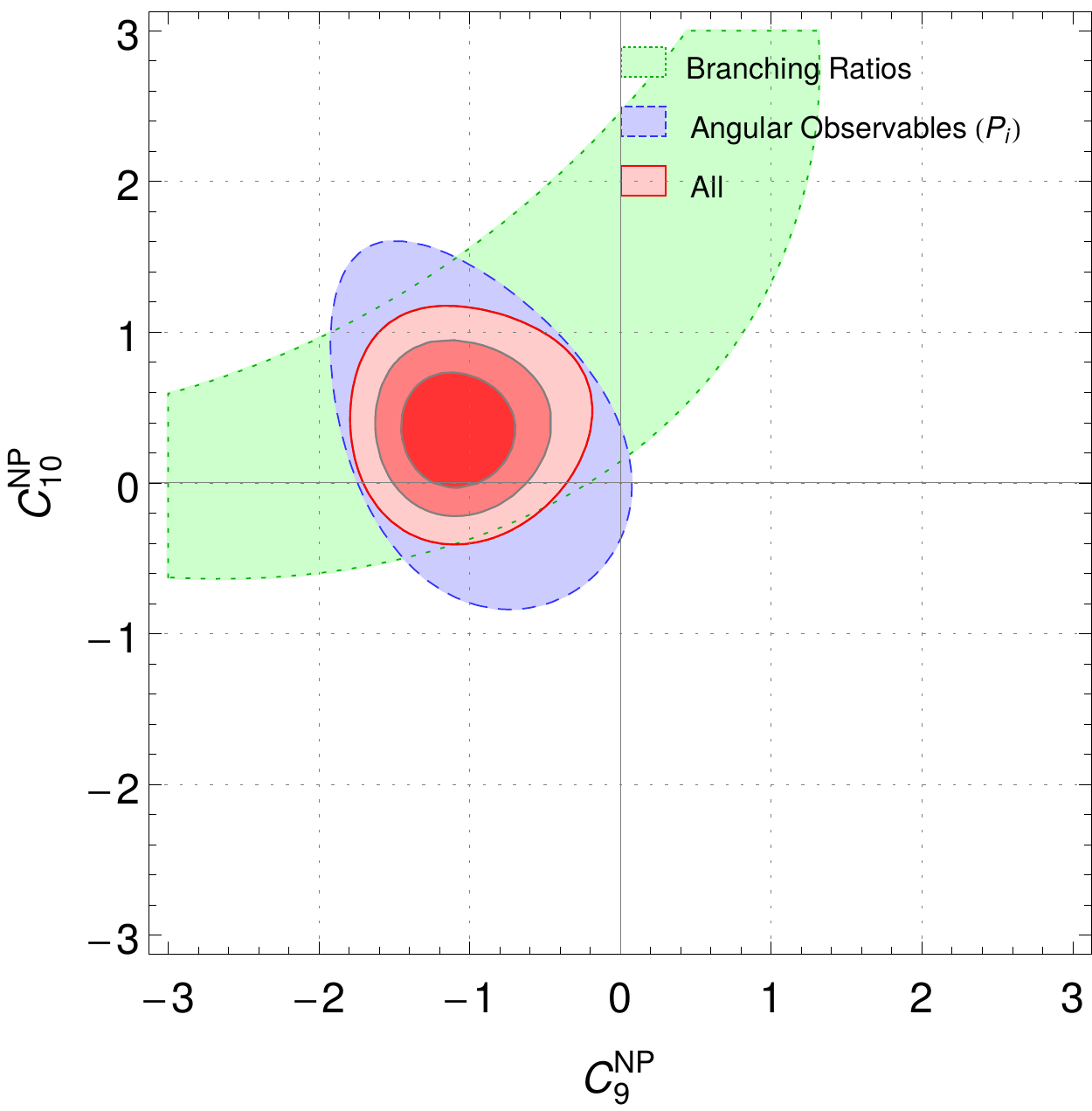}
\includegraphics[width=0.49\textwidth]{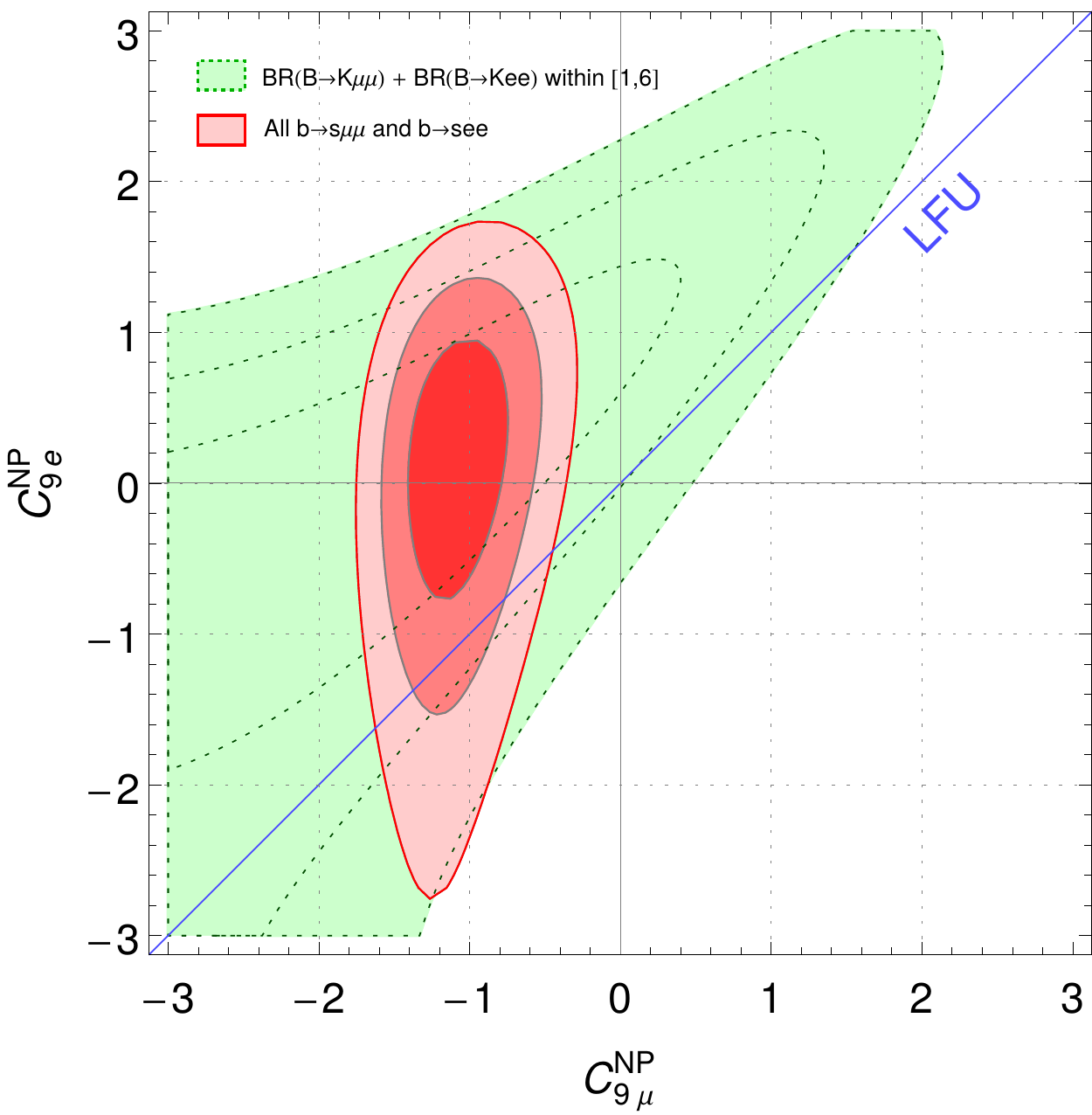}
\caption{Global fits of $b\to s\ell\ell$ anomalies in the $(C_9^{\rm NP},C_{10}^{\rm NP})$ (left) and $(C_{9\mu}^{\rm NP},C_{9e})^{\rm NP}$ (right) planes. The plots are taken from Ref.~\cite{Descotes-Genon:2015uva}.} 
\label{Fig:global}
\end{figure}

In the two panels of Fig.~\ref{Fig:global}, taken from Ref.~\cite{Descotes-Genon:2015uva}, we show the results of global fits in the $(C_9^{\rm NP},C_{10}^{\rm NP})$ plane assuming lepton flavour universality and in the $(C_{9\mu}^{\rm NP},C_{9e}^{\rm NP})$ plane after the inclusion of constraints from $R_K = {\rm BR} (B\to K\mu\mu)/{\rm BR} (B\to Kee)$ (see Ref.~\cite{Bordone:2016gaq} for a review of the theoretical uncertainties on $R_K$).  

New physics in $B\to K^{(*)}\ell\ell$ decays can be tested in the corresponding inclusive mode $B\to X_s\ell\ell$ for which power corrections are under much better theoretical control (see Ref.~\cite{Huber:2015sra} for a theoretical review). Issues related to the calculation of the $X_s$ invariant-mass spectrum, which is relevant in order to asses the impact of required experimental cuts on $m_{X_s}$, have been studied within a Fermi motion model~\cite{Ali:1996bm} and SCET~\cite{Lee:2005pwa, Benzke:2017}. Current theoretical predictions have uncertainties that are at the level of the experimental accuracy achievable by Belle~II with 50~${\rm fb}^{-1}$ of integrated luminosity~\cite{Hurth:2014vma}.

%% file: rarec.tex
\section{Rare $D$ decays}

Rare decays of $D$ mesons test for new physics in the flavour-changing
neutral currents involving up-type quarks. In the SM, the short-distance
contributions are strongly GIM suppressed and long-distance effects are
expected to dominate. This makes it very challenging to disentangle
potential new-physics contributions from the SM background. Correlations
between different measurements can help to discriminate the effects of new
physics.

As an example, in the SM, short-distance physics
contributes about $10^{-18}$ to ${\rm BR}(D\to\mu\mu)$, while the total BR is
dominated by the long-distance amplitude mediated by two
photons~\cite{Burdman:2001tf}. Specifically, ${\rm BR}(D\to\mu\mu)$ is expected
to be about $2.7\times10^{-5}\times {\rm BR}(D\to\gamma\gamma)$, or about
$10^{-13}$, with ${\rm BR}(D\to\gamma\gamma)$ expected to be $\leq 10^{-8}$.
In various new-physics models, ${\rm BR}(D\to\mu\mu)$ is correlated with
the value of the mixing parameter $x_D$. For example, in theories with
heavy vector-like quarks, FCNC interactions are generated in
the left-handed up-quark sector~\cite{Golowich:2009ii}, leading to the
prediction
${\rm BR}(D\to\mu\mu) \approx 4.3\times10^{-9}x_D \leq 4.3\times10^{-11}$.

For the $D\to\gamma\gamma$ decay,
the short-distance, EM-penguin contribution to the BR is
$3\times10^{-11}$~\cite{Greub:1996wn}, while the long-distance
contribution from vector-meson dominance is
(1--3)~$\times10^{-8}$~\cite{Fajfer:2001ad}.
A new-physics contribution from the MSSM (due to
gluino exchange via $c\to u\gamma$ transitions) would give $6\times10^{-6}$
\cite{Paul:2010pq}.
The most stringent limit, recently published by Belle, was obtained
from $D^{*+}\to D_0\pi^+$: ${\rm BR}(D\to\gamma\gamma) < 8.5\times10^{-7}$
(90\%~CL) \cite{Nisar:2015gvd}. Note that this provides an experimental
upper bound for the expected value of the long-distance SM contribution
to ${\rm BR}(D\to\mu\mu)$.

For the decays $D\to P\ell\ell$, the GIM
suppression is very effective and the SM rates are expected to be small
and dominated by long-distance effects, including production of the
$\rho/\omega$ and $\phi$ resonances, which contribute to the BR at the level
of $10^{-6}$. The non-resonant contribution to the BR is at the level of
$10^{-12}$~\cite{Fajfer:2007dy,deBoer:2015boa}.
New physics could contribute to the BR at the level of $10^{-9}$ in the
case of heavy vector-like quarks in the up sector, or as much as a few
$\times 10^{-6}$ in the case of the MSSM with $R$-parity
violation. The effects could be obscured by the
resonant contribution, but measurement of the decay spectrum
$d{\rm BR}/dq^2$ could help to isolate contributions from
new physics~\cite{Fajfer:2007dy}.
BES-III has recent preliminary results for the decays $D^+\to K^\pm e^+ e^\mp$
and $D^+\to\pi^\pm e^+e^\mp$, obtained with single-tagged events from
$\psi(3770)$ decays. The observation of the lepton-number-violating decays
$D^+\to K^-(\pi^-)e^+e^+$ would indicate the existence
of a Majorana neutrino mass term~~\cite{Zhang:2010um}. The limits
from BES-III on the BRs for $K^-e^+e^+$ ($6\times10^{-7}$) and
$\pi^+e^+e^-$ ($3\times10^{-7}$) are the most stringent obtained to date for
these lepton-number-violating and -conserving decays, respectively.
BES-III also presented a new limit on $D^+\to D^0e^+\nu_e$ at this conference:
${\rm BR} < 8.7\times10^{-5}$ (90\% CL), obtained via a double-tag technique
with the $D^0$ decaying to $K\pi$, $K\pi\pi^0$, or $K\pi\pi\pi$. The expected
BR in the SU(3) limit is $2.78\times10^{-13}$~\cite{Li:2007kgb}.

The short-distance contributions for the decays
$D^*(B^*)\to\ell\ell$~\cite{Khodjamirian:2015dda} are
expected to be larger than the long-distance contributions. Since the
SM BRs are on the order of $10^{-19}$ they cannot be measured directly.
Various new-physics models could increase the rate by orders of magnitude.
For instance, in the presence of a flavour-changing $Z'$ coupling to
left-handed quark currents, the BR could be as high as $2.5\times10^{-11}$.
It may be possible to measure the resonant production of
$D^*(2007)$ and subsequent strong or electromagnetic decay to $D^0\pi^0$ or
$D^0\gamma$ in $e^+e^-$ collisions at
$\sqrt{s} = m_{D^*}$,
for example, at BEPC-II with the BES-III detector, or possibly at VEPP-2000
with an upgrade to increase the machine energy above 2~GeV.

$D$ (or $B$) decays with missing energy could be a dark-matter
signature~\cite{Badin:2010uh}.
The $D\to\nu\bar{\nu}$ decay is helicity suppressed and has an SM BR of
$1.1\times10^{-30}$. Adding a final-state photon overcomes
the helicity suppression, but the SM BR is still $3.96\times10^{-14}$.
In a minimal scalar dark matter model, the BR for $D\to SS$ (with $S$ a
scalar dark matter particle) could be
on the order of $10^{-14}$.
Belle has recently published the first limit on
${\rm BR}(D^0\to{\rm invisible})$ obtained using the charm-tag
method, in which $e^+e^-\to D^{(*)}_{\rm tag} X_{\rm frag} \bar{D}^{*-}_{\rm sig}$,
$D^{(*)}_{\rm tag}$ and $X_{\rm frag}$ are reconstructed from a variety of modes,
and the $\pi^-$ from  $\bar{D}^{*-}_{\rm sig}\to\bar{D}^0\pi^-$ is
reconstructed as well. The result is ${\rm BR} < 8.8\times10^{-5}$
(90\% CL)~\cite{Lai:2016uvj}.

%% file: rarek.tex
\section{Rare $K$ decays}

The SM calculation of the BRs for the $K\to\pi\nu\bar{\nu}$ decays
is particularly clean because the loop amplitudes are dominated by
the top-quark contributions, the hadronic matrix element can be obtained
from the precisely known $K_{e3}$ rate, and there are no long-distance
contributions from processes with intermediate photons. 
In the SM,
${\rm BR}(K^+\to\pi^+\nu\bar{\nu}) = (8.4 \pm 1.0)\times10^{-11}$ and
${\rm BR}(K^0_L\to\pi^0\nu\bar{\nu}) = (3.4 \pm 0.6)\times10^{-11}$
\cite{Buras:2015qea}.
The dominant uncertainties are from $V_{cb}$, $V_{ub}$ and $\gamma$; the
underlying theoretical uncertainties are just $0.30$ and $0.05\times10^{-11}$,
respectively. If both BRs are measured and one or both is different
from its SM value, it may be possible to characterize the new physics
responsible (see e.g. \cite{Buras:2015yca}).

Seven candidate $K^+\to\pi^+\nu\bar{\nu}$ events have been seen by 
Brookhaven experiment E787 and its successor, E949, giving
${\rm BR}(K^+\to\pi^+\nu\bar{\nu}) =  1.73^{+1.15}_{-1.05}\times10^{-10}$
\cite{Artamonov:2009sz}. The goal of the NA62 experiment~\cite{NA62:2017rwk} is
to measure ${\rm BR}(K^+\to\pi^+\nu\bar{\nu})$ with a precision of about 10\%.
NA62 makes use of a 750~MHz, 75-GeV positive secondary beam from the CERN SPS,
providing 5~MHz of $K^+$ decays in a 60-m fiducial volume. The full beam rate
is tracked through three stations of silicon pixel detectors, and kaons in
the beam are identified by a differential Cerenkov counter. Secondaries are
tracked through four ultra-light straw tube stations operated in vacuum.
Precise reconstruction of the missing mass at the decay vertex provides
rejection against the dominant decays $K^+\to\mu^+\nu$ and $K^+\to\pi^+\pi^0$.
Photons from $\pi^0$ are vetoed by EM calorimeters with hermetic coverage for
angles below 50~mrad, including the NA48 liquid-krypton calorimeter (LKr).
Muon vetoes and hadron calorimeters downstream of the LKr provide
$\mu/\pi$ separation, while a RICH provides additional particle identification
for secondaries and helps with the precision time measurement
needed for accurate $K^+$-$\pi^+$ association.
NA62 is currently taking data and collected $10^{12}$ $K^+$ decays in 2016.
The experiment aims to collect $10^{13}$ $K^+$ decays to observe $\sim100$
signal events by the end of 2018.
NA62 can collect several triggers simultaneously to address a
broad physics portfolio, and plans to explore the dark sector in runs
after LS2.

The decay $K^0_L\to\pi^0\nu\bar{\nu}$ has never been observed.
Model-independent arguments~\cite{Grossman:1997sk} and the measured value of
${\rm BR}(K^+\to\pi^+\nu\bar{\nu})$
lead to the Grossman-Nir upper limit
${\rm BR}(K^0_L\to\pi^0\nu\bar{\nu}) < 1.4\times10^{-9}$.
The KOTO experiment at J-PARC (Tokai, Japan) is making steady progress at
pushing down the direct limit.
The experiment makes use of
a tightly collimated neutral beam (average momentum 2 GeV) and
compact, hermetic
detector. From a brief pilot run in 2013, KOTO obtained the limit
${\rm BR}(K^0_L\to\pi^0\nu\bar{\nu}) < 5.1\times10^{-8}$
(90\% CL)~\cite{Ahn:2016kja}. The experiment ran for many months in
2015--2016, with various improvements to reduce the background.
From the preliminary analysis of 10\% of the data collected in 2015,
KOTO has reached a single-event BR sensitivity of $5.9\times10^{-9}$, with
background levels still under evaluation.
With the entire 2015--2016 data set, KOTO should be able to push the
single-event sensitivity to below the Grossman-Nir bound.
By the end of 2015, the beam power reached 42~kW; it is expected to
increase to 100~kW by 2018.
A series of upgrades to the experiment are also in progress,
including an additional layer for the barrel calorimeter
(tested in 2016) and front/back readout for the
forward CsI calorimeter to provide additional neutron rejection.
The experiment expects to reach single-event sensitivity at the level of
the SM BR by about 2021.

Numerous other rare kaon decays are of interest besides
$K\to\pi\nu\bar{\nu}$.
Since the $CP$-violating phase of the $s\to d\ell^+\ell^-$ transition
is poorly determined, there is room for new physics to manifest itself
in $K\to\pi\ell^+\ell^-$ and $K\to\ell^+\ell^-$ decays, although there
are complications from the long-distance contributions~\cite{Smith:2014mla}.
For example, $K_L\to\mu^+\mu^-$ is dominated by the long-distance component
arising via $K_L\to\gamma\gamma$.
On the other hand, about one-third of the $K_S\to\mu^+\mu^-$ width
(${\rm BR}\sim5\times10^{-12}$) is from the $CP$-violating short-distance
amplitude~\cite{Isidori:2003ts}. The recent limit from LHCb,
${\rm BR}(K_S\to\mu^+\mu^-) < 5.8\times10^{-9}$ (90\%~CL) \cite{LHCb:2016meg},
demonstrates LHCb's capability to measure $K_S$ decays to muons.
LHCb may also be able to measure ${\rm BR}(K_S\to\pi^0\mu^+\mu^-)$, which
helps to isolate the short-distance component of $K_L\to\pi^0\mu^+\mu^-$ by
pinning down the contribution from indirect $CP$ violation.
The $K^+\to\pi^+\ell^+\ell^-$ decays may show signs of lepton-flavour
universality violation, in analogy to the $R_K$ and $P'_5$ anomalies
observed in the $B$ system. NA62 may have the statistics to considerably
improve on NA48/2 results for the form-factor slopes for these decays.
High-statistics measurements of the Dalitz plots for the
$K\to\pi\pi\gamma$ and $K\to\pi\pi ee$ decays would allow the
inner-bremsstrahlung and direct-emission contributions to be isolated,
enabling searches for $CP$-violating amplitudes in the latter case.
Finally, NA62 and other experiments measuring $\pi^\pm$
and $K^\pm$ decays could also look for time-dependent anisotropies in the
lepton momentum distribution in $P_{\mu2}$, which would be observable
if Lorentz invariance were violated and the fundamental symmetry group
were instead the SIM(2) subgroup, implying the existence of a preferred
direction in spacetime~\cite{Nayak:2016zed}.

%% file: chargeLFV.tex
\section{Charged lepton flavour violation}

In the SM,  charged lepton flavour violating decays are highly suppressed due to the smallness of neutrino masses. 
The observation of such a decay would therefore be an unambiguous sign of physics beyond the SM.

Lepton flavour violating decays of muons are searched for at dedicated facilities. The MEG experiment at PSI has improved the limit on the $\mu^+\to e^+ \gamma$ decay by a factor 30 with respect to the previous experiment using their final data set:  ${\rm BR}(\mu^+\to e^+ \gamma)< 4.2\times 10^{-13}$ at 90\% CL \cite{TheMEG:2016wtm}. The experiment has been upgraded to reach a sensitivity of $4\times 10^{-14}$ and will have a first engineering run in 2017.
The  $\mu^+\to e^+ e^- e^+$ decay has been searched for previously by the SINDRUM experiment reaching,  $1.0\times 10^{-12}$. The Mu3e experiment at PSI aims at reaching a sensitivity of $10^{-15}$ as a first step in  2018--2020, and  $10^{-16}$ in a second step. Muon-to-electron conversion
will be searched for by COMET at J-PARC and Mu2e at Fermilab. 
They both expect to reach a single event sensitivity of $\sim 2\times10^{-17}$ using aluminium targets in the years 2020.

Lepton flavour violation is also searched for in $\tau$ and $B$ decays at the $B$ factories and LHC experiments. 
A few new results have appeared in the last couple of years. The majority of the limits for $\tau$ decays are dominated by the Belle collaboration,  reaching limits in the range $10^{-8}$ to $2\times10^{-7}$,  except for $\tau^+ \to e^+/\mu^+ \gamma$, where Babar has the most stringent limits, and $\tau \to p \mu^+ \mu^- $, which has only been searched for by LHCb. 
Regarding the $B$ decays, Babar has obtained the best limits in most cases, spanning a range from a few~$\times10^{-8}$ for the $e$ and $\mu$ channels
to a few~$\times10^{-5}$ for channels involving $\tau$ leptons. LHCb has obtained the best limits for the $B^0\to e^\pm \mu^\mp$ channel, reaching a limit at $2.8\times 10^{-9}$ (90\% CL) with 1 $\mathrm{fb}^{-1}$ of Run 1 data.
For a complete picture of $\tau$ and $B$ LFV results, see Ref.~\cite{Amhis:2016xyh}.

%% file: summary.tex
\section{Summary}

While generally speaking, present results on rare $B$ decays are in agreement
with SM predictions, measurements of some observables, notably $R_K$ and
$P'_5$, show intriguing hints of significant deviation. It is very encouraging
to see the excellent reach of LHCb with only Run 1 data, as well as the
improvements in the results from the $B$-factory experiments, which are
continuously refining their analysis techniques to obtain more sensitive
tests of the SM. This year's CKM conference has also seen the first results
on rare $B$ decays to $\tau$ leptons, which are interesting in view of the
hints of lepton flavour universality violation in decays to muons.
Within the next few years, new results from LHCb, ATLAS, and CMS Run 2
data will be supplemented by the arrival of the first results from
Belle II and from the kaon experiments KOTO and NA62, as well as by results
from the next generation of charged lepton flavour violation experiments.
The prospects are therefore excellent for rare decays to continue to place
new and increasingly tight constraints on the Standard Model.

%% file: ckm16-wg3.bbl
\begin{thebibliography}{99}


\bibitem{Cabibbo:1963yz} 
  N.~Cabibbo,
  %``Unitary Symmetry and Leptonic Decays,''
  Phys.\ Rev.\ Lett.\  {\bf 10}, 531 (1963).
  doi:10.1103/PhysRevLett.10.531
  %%CITATION = doi:10.1103/PhysRevLett.10.531;%%


\bibitem{Kobayashi:1973fv} 
  M.~Kobayashi and T.~Maskawa,
  %``CP Violation in the Renormalizable Theory of Weak Interaction,''
  Prog.\ Theor.\ Phys.\  {\bf 49}, 652 (1973).
  doi:10.1143/PTP.49.652
  %%CITATION = doi:10.1143/PTP.49.652;%%

\bibitem{belleII}
Talk by R. Itoh at "Sixth Workshop on Theory, Phenomenology and Experiments in Flavour Physics - FPCapri2016", Capri, Italy, 11-13 June 2016.

\bibitem{NA62:2017rwk}
  E.~Cortina Gil {\it et al.} [NA62 Collaboration],
  %``The Beam and detector of the NA62 experiment at CERN,''
  arXiv:1703.08501 [physics.ins-det].
  %%CITATION = ARXIV:1703.08501;%%
  %1 citations counted in INSPIRE as of 14 Apr 2017

\bibitem{Komatsubara:2012pn} 
  T.~K.~Komatsubara,
  %``Experiments with K-Meson Decays,''
  Prog.\ Part.\ Nucl.\ Phys.\  {\bf 67}, 995 (2012)
  doi:10.1016/j.ppnp.2012.04.001
  [arXiv:1203.6437 [hep-ex]].
  %%CITATION = doi:10.1016/j.ppnp.2012.04.001;%%

\bibitem{Ahn:2016kja}
  J.~K.~Ahn {\it et al.},
  %``A new search for the $K_{L} \to \pi^0 \nu \overline{\nu}$ and $K_{L} \to \pi^{0} X^{0}$ decays,''
  arXiv:1609.03637 [hep-ex].
  %%CITATION = ARXIV:1609.03637;%%
  %5 citations counted in INSPIRE as of 16 Mar 2017

%%%%%%% leptonic

%\cite{Bobeth:2013uxa}
\bibitem{Bobeth:2013uxa}
  C.~Bobeth, M.~Gorbahn, T.~Hermann, M.~Misiak, E.~Stamou and M.~Steinhauser,
  %``$B_{s,d} \to l^+ l^-$ in the Standard Model with Reduced Theoretical Uncertainty,''
  Phys.\ Rev.\ Lett.\  {\bf 112} (2014) 101801
  doi:10.1103/PhysRevLett.112.101801
  [arXiv:1311.0903 [hep-ph]].
  %%CITATION = doi:10.1103/PhysRevLett.112.101801;%%
  %203 citations counted in INSPIRE as of 10 Mar 2017

%\cite{Aaboud:2016ire}
\bibitem{Aaboud:2016ire}
  M.~Aaboud {\it et al.} [ATLAS Collaboration],
  %``Study of the rare decays of $B^0_s$ and $B^0$ into muon pairs from data collected during the LHC Run 1 with the ATLAS detector,''
  Eur.\ Phys.\ J.\ C {\bf 76} (2016) no.9,  513
  doi:10.1140/epjc/s10052-016-4338-8
  [arXiv:1604.04263 [hep-ex]].
  %%CITATION = doi:10.1140/epjc/s10052-016-4338-8;%%
  %26 citations counted in INSPIRE as of 10 Mar 2017

%\cite{DeBruyn:2016tiq}
\bibitem{DeBruyn:2016tiq}
  K.~De Bruyn [LHCb Collaboration],
  %``Search for the rare decays $B^0_{(s)}\to\tau^+\tau^-$,''
  LHCb-CONF-2016-011, CERN-LHCb-CONF-2016-011.
  %%CITATION = LHCB-CONF-2016-011, CERN-LHCB-CONF-2016-011;%%
  %1 citations counted in INSPIRE as of 10 Mar 2017

%%%%% semileptonic

%\cite{Aaij:2014ora}
\bibitem{Aaij:2014ora}
  R.~Aaij {\it et al.} [LHCb Collaboration],
  %``Test of lepton universality using $B^{+}\rightarrow K^{+}\ell^{+}\ell^{-}$ decays,''
  Phys.\ Rev.\ Lett.\  {\bf 113} (2014) 151601
  doi:10.1103/PhysRevLett.113.151601
  [arXiv:1406.6482 [hep-ex]].
  %%CITATION = doi:10.1103/PhysRevLett.113.151601;%%
  %292 citations counted in INSPIRE as of 10 Mar 2017

%\cite{Bordone:2016gaq}
\bibitem{Bordone:2016gaq}
  M.~Bordone, G.~Isidori and A.~Pattori,
  %``On the Standard Model predictions for $R_K$ and $R_{K^*}$,''
  Eur.\ Phys.\ J.\ C {\bf 76} (2016) no.8,  440
  doi:10.1140/epjc/s10052-016-4274-7
  [arXiv:1605.07633 [hep-ph]].
  %%CITATION = doi:10.1140/epjc/s10052-016-4274-7;%%
  %25 citations counted in INSPIRE as of 10 Mar 2017

%\cite{Aaij:2014pli}
\bibitem{Aaij:2014pli}
  R.~Aaij {\it et al.} [LHCb Collaboration],
  %``Differential branching fractions and isospin asymmetries of $B \to K^{(*)} \mu^+ \mu^-$ decays,''
  JHEP {\bf 1406} (2014) 133
  doi:10.1007/JHEP06(2014)133
  [arXiv:1403.8044 [hep-ex]].
  %%CITATION = doi:10.1007/JHEP06(2014)133;%%
  %118 citations counted in INSPIRE as of 10 Mar 2017

%\cite{Aaij:2015xza}
\bibitem{Aaij:2015xza}
  R.~Aaij {\it et al.} [LHCb Collaboration],
  %``Differential branching fraction and angular analysis of $\Lambda^{0}_{b} \rightarrow \Lambda \mu^+\mu^-$ decays,''
  JHEP {\bf 1506} (2015) 115
  doi:10.1007/JHEP06(2015)115
  [arXiv:1503.07138 [hep-ex]].
  %%CITATION = doi:10.1007/JHEP06(2015)115;%%
  %31 citations counted in INSPIRE as of 10 Mar 2017


%\cite{Aaij:2015esa}
\bibitem{Aaij:2015esa}
  R.~Aaij {\it et al.} [LHCb Collaboration],
  %``Angular analysis and differential branching fraction of the decay $B^0_s\to\phi\mu^+\mu^-$,''
  JHEP {\bf 1509} (2015) 179
  doi:10.1007/JHEP09(2015)179
  [arXiv:1506.08777 [hep-ex]].
  %%CITATION = doi:10.1007/JHEP09(2015)179;%%
  %62 citations counted in INSPIRE as of 10 Mar 2017

%\cite{Aaij:2016flj}
\bibitem{Aaij:2016flj}
  R.~Aaij {\it et al.} [LHCb Collaboration],
  %``Measurements of the S-wave fraction in $B^{0}\rightarrow K^{+}\pi^{-}\mu^{+}\mu^{-}$ decays and the $B^{0}\rightarrow K^{\ast}(892)^{0}\mu^{+}\mu^{-}$ differential branching fraction,''
  JHEP {\bf 1611} (2016) 047
  doi:10.1007/JHEP11(2016)047
  [arXiv:1606.04731 [hep-ex]].
  %%CITATION = doi:10.1007/JHEP11(2016)047;%%
  %8 citations counted in INSPIRE as of 10 Mar 2017
  
%\cite{Lees:2015ymt}
\bibitem{Lees:2015ymt}
  J.~P.~Lees {\it et al.} [BaBar Collaboration],
  %``Measurement of angular asymmetries in the decays $B \to K^*ℓ^+ℓ^-$,''
  Phys.\ Rev.\ D {\bf 93} (2016) no.5,  052015
  doi:10.1103/PhysRevD.93.052015
  [arXiv:1508.07960 [hep-ex]].
  %%CITATION = doi:10.1103/PhysRevD.93.052015;%%
  %11 citations counted in INSPIRE as of 10 Mar 2017

%\cite{Khachatryan:2015isa}
\bibitem{Khachatryan:2015isa}
  V.~Khachatryan {\it et al.} [CMS Collaboration],
  %``Angular analysis of the decay $B^0 \to K^{*0} \mu^+ \mu^-$ from pp collisions at $\sqrt  s = 8$ TeV,''
  Phys.\ Lett.\ B {\bf 753} (2016) 424
  doi:10.1016/j.physletb.2015.12.020
  [arXiv:1507.08126 [hep-ex]].
  %%CITATION = doi:10.1016/j.physletb.2015.12.020;%%
  %21 citations counted in INSPIRE as of 10 Mar 2017

%\cite{Aaij:2015oid}
\bibitem{Aaij:2015oid}
  R.~Aaij {\it et al.} [LHCb Collaboration],
  %``Angular analysis of the $B^{0} \to K^{*0} \mu^{+} \mu^{-}$ decay using 3 fb$^{-1}$ of integrated luminosity,''
  JHEP {\bf 1602} (2016) 104
  doi:10.1007/JHEP02(2016)104
  [arXiv:1512.04442 [hep-ex]].
  %%CITATION = doi:10.1007/JHEP02(2016)104;%%
  %107 citations counted in INSPIRE as of 10 Mar 2017

%\cite{Wehle:2016yoi}
\bibitem{Wehle:2016yoi}
  S.~Wehle {\it et al.} [Belle Collaboration],
  %``Lepton-Flavor-Dependent Angular Analysis of $B\to K^\ast \ell^+\ell^-$,''
  arXiv:1612.05014 [hep-ex].
  %%CITATION = ARXIV:1612.05014;%%
  %7 citations counted in INSPIRE as of 10 Mar 2017
  
 %\cite{Descotes-Genon:2014uoa}
 \bibitem{Descotes-Genon:2014uoa}
   S.~Descotes-Genon, L.~Hofer, J.~Matias and J.~Virto,
   %``On the impact of power corrections in the prediction of $B \to K^*\mu^+\mu^-$ observables,''
   JHEP {\bf 1412} (2014) 125
   doi:10.1007/JHEP12(2014)125
   [arXiv:1407.8526 [hep-ph]].
   %%CITATION = doi:10.1007/JHEP12(2014)125;%%
   %113 citations counted in INSPIRE as of 10 Mar 2017

%\cite{Capdevila:2016ivx}
\bibitem{Capdevila:2016ivx}
  B.~Capdevila, S.~Descotes-Genon, J.~Matias and J.~Virto,
  %``Assessing lepton-flavour non-universality from $B\to K^*\ell\ell$ angular analyses,''
  JHEP {\bf 1610} (2016) 075
  doi:10.1007/JHEP10(2016)075
  [arXiv:1605.03156 [hep-ph]].
  %%CITATION = doi:10.1007/JHEP10(2016)075;%%
  %14 citations counted in INSPIRE as of 10 Mar 2017

\bibitem{Bhatia:2017tgo}
  D.~Bhatia, S.~Chakraborty and A.~Dighe,
  %``Neutrino mixing and $R_K$ anomaly in $U(1)_X$ models: a bottom-up approach,''
  arXiv:1701.05825 [hep-ph].
  %%CITATION = ARXIV:1701.05825;%%

%\cite{Hambrock:2015wka}
\bibitem{Hambrock:2015wka}
  C.~Hambrock, A.~Khodjamirian and A.~Rusov,
  %``Hadronic effects and observables in $B\to \pi\ell^{+}\ell^{-}$ decay at large recoil,''
  Phys.\ Rev.\ D {\bf 92} (2015) no.7,  074020
  doi:10.1103/PhysRevD.92.074020
  [arXiv:1506.07760 [hep-ph]].
  %%CITATION = doi:10.1103/PhysRevD.92.074020;%%
  %10 citations counted in INSPIRE as of 10 Mar 2017
%\cite{Buras:2014fpa}
\bibitem{Buras:2014fpa}
  A.~J.~Buras, J.~Girrbach-Noe, C.~Niehoff and D.~M.~Straub,
  %``$ B\to {K}^{\left(\ast \right)}\nu \overline{\nu} $ decays in the Standard Model and beyond,''
  JHEP {\bf 1502} (2015) 184
  doi:10.1007/JHEP02(2015)184
  [arXiv:1409.4557 [hep-ph]].
  %%CITATION = doi:10.1007/JHEP02(2015)184;%%
  %68 citations counted in INSPIRE as of 10 Mar 2017

%\cite{Lutz:2013ftz}
\bibitem{Lutz:2013ftz}
  O.~Lutz {\it et al.} [Belle Collaboration],
  %``Search for $B \to h^{(*)} \nu \bar{\nu}$ with the full Belle $\Upsilon(4S)$ data sample,''
  Phys.\ Rev.\ D {\bf 87} (2013) no.11,  111103
  doi:10.1103/PhysRevD.87.111103
  [arXiv:1303.3719 [hep-ex]].
  %%CITATION = doi:10.1103/PhysRevD.87.111103;%%
  %47 citations counted in INSPIRE as of 10 Mar 2017

%\cite{Lees:2013kla}
\bibitem{Lees:2013kla}
  J.~P.~Lees {\it et al.} [BaBar Collaboration],
  %``Search for $B \to K^{(*)} \nu \overline \nu$ and invisible quarkonium decays,''
  Phys.\ Rev.\ D {\bf 87} (2013) no.11,  112005
  doi:10.1103/PhysRevD.87.112005
  [arXiv:1303.7465 [hep-ex]].
  %%CITATION = doi:10.1103/PhysRevD.87.112005;%%
  %52 citations counted in INSPIRE as of 10 Mar 2017

%\cite{delAmoSanchez:2010bk}
\bibitem{delAmoSanchez:2010bk}
  P.~del Amo Sanchez {\it et al.} [BaBar Collaboration],
  %``Search for the Rare Decay $B \to K \nu \bar{\nu}$,''
  Phys.\ Rev.\ D {\bf 82} (2010) 112002
  doi:10.1103/PhysRevD.82.112002
  [arXiv:1009.1529 [hep-ex]].
  %%CITATION = doi:10.1103/PhysRevD.82.112002;%%
  %35 citations counted in INSPIRE as of 10 Mar 2017

%\cite{Muheim:2008vu}
\bibitem{Muheim:2008vu} 
  F.~Muheim, Y.~Xie and R.~Zwicky,
  %``Exploiting the width difference in $B_s \to \phi \gamma$,''
  Phys.\ Lett.\ B {\bf 664}, 174 (2008)
  doi:10.1016/j.physletb.2008.05.032
  [arXiv:0802.0876 [hep-ph]].
  %%CITATION = doi:10.1016/j.physletb.2008.05.032;%%
  %59 citations counted in INSPIRE as of 24 Apr 2017

%\cite{Aaij:2016ofv}
\bibitem{Aaij:2016ofv} 
  R.~Aaij {\it et al.} [LHCb Collaboration],
  %``First experimental study of photon polarization in radiative $B^{0}_{s}$ decays,''
  Phys.\ Rev.\ Lett.\  {\bf 118}, no. 2, 021801 (2017)
  Addendum: [Phys.\ Rev.\ Lett.\  {\bf 118}, no. 10, 109901 (2017)]
  doi:10.1103/PhysRevLett.118.021801, 10.1103/PhysRevLett.118.109901
  [arXiv:1609.02032 [hep-ex]].
  %%CITATION = doi:10.1103/PhysRevLett.118.021801, 10.1103/PhysRevLett.118.109901;%%
  %4 citations counted in INSPIRE as of 24 Apr 2017

  
%\cite{Grygier:2017tzo}
\bibitem{Grygier:2017tzo}
  J.~Grygier {\it et al.} [Belle Collaboration],
  %``Search for $\boldsymbol{B\to h\nu\bar{\nu}}$ decays with semileptonic tagging at Belle,''
  arXiv:1702.03224 [hep-ex].
  %%CITATION = ARXIV:1702.03224;%%
  %2 citations counted in INSPIRE as of 10 Mar 2017






%\cite{Aaij:2017ewm}
\bibitem{Aaij:2017ewm}
  R.~Aaij {\it et al.} [LHCb Collaboration],
  %``Observation of the suppressed decay $\Lambda^{0}_{b}\rightarrow p\pi^{-}\mu^{+}\mu^{-}$,''
  arXiv:1701.08705 [hep-ex].
  %%CITATION = ARXIV:1701.08705;%%

%\cite{Detmold:2016pkz}
\bibitem{Detmold:2016pkz}
  W.~Detmold and S.~Meinel,
  %``$\Lambda_b \to \Lambda \ell^+ \ell^-$ form factors, differential branching fraction, and angular observables from lattice QCD with relativistic $b$ quarks,''
  Phys.\ Rev.\ D {\bf 93} (2016) no.7,  074501
  doi:10.1103/PhysRevD.93.074501
  [arXiv:1602.01399 [hep-lat]].
  %%CITATION = doi:10.1103/PhysRevD.93.074501;%%


%\cite{TheBaBar:2016xwe}
\bibitem{TheBaBar:2016xwe}
  J.~P.~Lees {\it et al.} [BaBar Collaboration],
  %``Search for $B^{+}\rightarrow K^{+} \tau^{+}\tau^{-}$ at the BaBar experiment,''
  Phys.\ Rev.\ Lett.\  {\bf 118} (2017) no.3,  031802
  doi:10.1103/PhysRevLett.118.031802
  [arXiv:1605.09637 [hep-ex]].
  %%CITATION = doi:10.1103/PhysRevLett.118.031802;%%
  %8 citations counted in INSPIRE as of 09 Mar 2017















%%%%%%% Radiative

\bibitem{Misiak:2015xwa} 
  M.~Misiak {\it et al.},
  %``Updated NNLO QCD predictions for the weak radiative B-meson decays,''
  Phys.\ Rev.\ Lett.\  {\bf 114}, no. 22, 221801 (2015)
  doi:10.1103/PhysRevLett.114.221801
  [arXiv:1503.01789 [hep-ph]].
  %%CITATION = doi:10.1103/PhysRevLett.114.221801;%%

\bibitem{Czakon:2015exa} 
  M.~Czakon, P.~Fiedler, T.~Huber, M.~Misiak, T.~Schutzmeier and M.~Steinhauser,
  %``The $(Q_{7}, Q_{1,2})$ contribution to $ \overline{B}\to {X}_s\gamma $ at $ \mathcal{O}\left({\alpha}_{\mathrm{s}}^2\right) $,''
  JHEP {\bf 1504}, 168 (2015)
  doi:10.1007/JHEP04(2015)168
  [arXiv:1503.01791 [hep-ph]].
  %%CITATION = doi:10.1007/JHEP04(2015)168;%%

\bibitem{Misiak:2017woa} 
  M.~Misiak, A.~Rehman and M.~Steinhauser,
  %``NNLO QCD counterterm contributions to B -> X_s gamma for the physical value of m_c,''
  arXiv:1702.07674 [hep-ph].
    %%CITATION = ARXIV:1702.07674;%%

\bibitem{Ewerth:2009yr} 
  T.~Ewerth, P.~Gambino and S.~Nandi,
  %``Power suppressed effects in anti-B ---> X(s) gamma at O(alpha(s)),''
  Nucl.\ Phys.\ B {\bf 830}, 278 (2010)
  doi:10.1016/j.nuclphysb.2009.12.035
  [arXiv:0911.2175 [hep-ph]].
  %%CITATION = doi:10.1016/j.nuclphysb.2009.12.035;%%
  
  \bibitem{Gambino:2016jkc} 
  P.~Gambino, K.~J.~Healey and S.~Turczyk,
  %``Taming the higher power corrections in semileptonic B decays,''
  Phys.\ Lett.\ B {\bf 763}, 60 (2016)
  doi:10.1016/j.physletb.2016.10.023
  [arXiv:1606.06174 [hep-ph]].
  %%CITATION = doi:10.1016/j.physletb.2016.10.023;%%
  
  \bibitem{Benzke:2010js} 
  M.~Benzke, S.~J.~Lee, M.~Neubert and G.~Paz,
  %``Factorization at Subleading Power and Irreducible Uncertainties in $\bar B\to X_s\gamma$ Decay,''
  JHEP {\bf 1008}, 099 (2010)
  doi:10.1007/JHEP08(2010)099
  [arXiv:1003.5012 [hep-ph]].
  %%CITATION = doi:10.1007/JHEP08(2010)099;%%
  
  
  \bibitem{Belle:2016ufb} 
  A.~Abdesselam {\it et al.} [Belle Collaboration],
  %``Measurement of the inclusive $B\to X_{s+d} \gamma$ branching fraction, photon energy spectrum and HQE parameters,''
  arXiv:1608.02344 [hep-ex].
  %%CITATION = ARXIV:1608.02344;%%
  
  \bibitem{Lyon:2013gba} 
  J.~Lyon and R.~Zwicky,
  %``Isospin asymmetries in $B\to(K^*,\rho)\gamma/l^+l^-$ and $B\to Kl^+l^-$ in and beyond the standard model,''
  Phys.\ Rev.\ D {\bf 88}, no. 9, 094004 (2013)
  doi:10.1103/PhysRevD.88.094004
  [arXiv:1305.4797 [hep-ph]].
  %%CITATION = doi:10.1103/PhysRevD.88.094004;%%
  
  \bibitem{Sanchez:2015pxu} 
  P.~del Amo Sanchez {\it et al.} [BaBar Collaboration],
  %``Time-dependent analysis of $B^0 \to {{K^0_{S}}} \pi^- \pi^+ \gamma$ decays and studies of the $K^+\pi^-\pi^+$ system in $B^+ \to K^+ \pi^- \pi^+ \gamma$ decays,''
  Phys.\ Rev.\ D {\bf 93}, no. 5, 052013 (2016)
  doi:10.1103/PhysRevD.93.052013
  [arXiv:1512.03579 [hep-ex]].
  %%CITATION = doi:10.1103/PhysRevD.93.052013;%%
  
  
  
  
  
  
  
  
  
  
  
  
  
 %%%%%%%%%% Global Fits 
  
\bibitem{Bauer:2000yr} 
  C.~W.~Bauer, S.~Fleming, D.~Pirjol and I.~W.~Stewart,
  %``An Effective field theory for collinear and soft gluons: Heavy to light decays,''
  Phys.\ Rev.\ D {\bf 63}, 114020 (2001)
  doi:10.1103/PhysRevD.63.114020
  [hep-ph/0011336].
  %%CITATION = doi:10.1103/PhysRevD.63.114020;%%  
  
  \bibitem{Beneke:2001at} 
  M.~Beneke, T.~Feldmann and D.~Seidel,
  %``Systematic approach to exclusive $B \to  V l^+ l^-$, $V \gamma$ decays,''
  Nucl.\ Phys.\ B {\bf 612}, 25 (2001)
  doi:10.1016/S0550-3213(01)00366-2
  [hep-ph/0106067].
  %%CITATION = doi:10.1016/S0550-3213(01)00366-2;%%

 
 \bibitem{Grinstein:2004vb} 
  B.~Grinstein and D.~Pirjol,
  %``Exclusive rare $B \to K^*\ell^+\ell^-$ decays at low recoil: Controlling the long-distance effects,''
  Phys.\ Rev.\ D {\bf 70}, 114005 (2004)
  doi:10.1103/PhysRevD.70.114005
  [hep-ph/0404250].
  %%CITATION = doi:10.1103/PhysRevD.70.114005;%%
 
  
  \bibitem{Bobeth:2010wg} 
  C.~Bobeth, G.~Hiller and D.~van Dyk,
  %``The Benefits of $\bar{B} -> \bar{K}^* l^+ l^-$ Decays at Low Recoil,''
  JHEP {\bf 1007}, 098 (2010)
  doi:10.1007/JHEP07(2010)098
  [arXiv:1006.5013 [hep-ph]].
  %%CITATION = doi:10.1007/JHEP07(2010)098;%%

 \bibitem{Du:2015tda} 
  D.~Du, A.~X.~El-Khadra, S.~Gottlieb, A.~S.~Kronfeld, J.~Laiho, E.~Lunghi, R.~S.~Van de Water and R.~Zhou,
  %``Phenomenology of semileptonic B-meson decays with form factors from lattice QCD,''
  Phys.\ Rev.\ D {\bf 93}, no. 3, 034005 (2016)
  doi:10.1103/PhysRevD.93.034005
  [arXiv:1510.02349 [hep-ph]].
  %%CITATION = doi:10.1103/PhysRevD.93.034005;%%
  
  \bibitem{Descotes-Genon:2015uva} 
  S.~Descotes-Genon, L.~Hofer, J.~Matias and J.~Virto,
  %``Global analysis of $b\to s\ell\ell$ anomalies,''
  JHEP {\bf 1606}, 092 (2016)
  doi:10.1007/JHEP06(2016)092
  [arXiv:1510.04239 [hep-ph]].
  %%CITATION = doi:10.1007/JHEP06(2016)092;%%

\bibitem{Jager:2012uw} 
  S.~J\"ager and J.~Martin Camalich,
  %``On $B \to  V \ell \ell$ at small dilepton invariant mass, power corrections, and new physics,''
  JHEP {\bf 1305}, 043 (2013)
  doi:10.1007/JHEP05(2013)043
  [arXiv:1212.2263 [hep-ph]].
  %%CITATION = doi:10.1007/JHEP05(2013)043;%%

\bibitem{Jager:2014rwa} 
  S.~J\"ager and J.~Martin Camalich,
  %``Reassessing the discovery potential of the $B \to K^{*} \ell^+\ell^-$ decays in the large-recoil region: SM challenges and BSM opportunities,''
  Phys.\ Rev.\ D {\bf 93}, no. 1, 014028 (2016)
  doi:10.1103/PhysRevD.93.014028
  [arXiv:1412.3183 [hep-ph]].
  %%CITATION = doi:10.1103/PhysRevD.93.014028;%%


\bibitem{Ciuchini:2015qxb} 
  M.~Ciuchini, M.~Fedele, E.~Franco, S.~Mishima, A.~Paul, L.~Silvestrini and M.~Valli,
  %``$B\to K^* \ell^+ \ell^-$ decays at large recoil in the Standard Model: a theoretical reappraisal,''
  JHEP {\bf 1606}, 116 (2016)
  doi:10.1007/JHEP06(2016)116
  [arXiv:1512.07157 [hep-ph]].
  %%CITATION = doi:10.1007/JHEP06(2016)116;%%

\bibitem{Beylich:2011aq} 
  M.~Beylich, G.~Buchalla and T.~Feldmann,
  %``Theory of $B \to K^{(*)}\ell^+ \ell^-$ decays at high $q^2$: OPE and quark-hadron duality,''
  Eur.\ Phys.\ J.\ C {\bf 71}, 1635 (2011)
  doi:10.1140/epjc/s10052-011-1635-0
  [arXiv:1101.5118 [hep-ph]].
  %%CITATION = doi:10.1140/epjc/s10052-011-1635-0;%%

  \bibitem{Lyon:2014hpa} 
  J.~Lyon and R.~Zwicky,
  %``Resonances gone topsy turvy - the charm of QCD or new physics in $b \to s \ell^+ \ell^-$?,''
  arXiv:1406.0566 [hep-ph].
  %%CITATION = ARXIV:1406.0566;%%



%B->pi and Bs->K RBC 
\bibitem{Flynn:2015mha} 
  J.~M.~Flynn, T.~Izubuchi, T.~Kawanai, C.~Lehner, A.~Soni, R.~S.~Van de Water and O.~Witzel,
  %``$B \to \pi \ell \nu$ and $B_s \to K \ell \nu$ form factors and $|V_{ub}|$ from 2+1-flavor lattice QCD with domain-wall light quarks and relativistic heavy quarks,''
  Phys.\ Rev.\ D {\bf 91}, no. 7, 074510 (2015)
  doi:10.1103/PhysRevD.91.074510
  [arXiv:1501.05373 [hep-lat]].
  %%CITATION = doi:10.1103/PhysRevD.91.074510;%%

%B->pi FNAL f+
\bibitem{Lattice:2015tia} 
  J.~A.~Bailey {\it et al.} [Fermilab Lattice and MILC Collaborations],
  %``$|V_{ub}|$ from $B\to\pi\ell\nu$ decays and (2+1)-flavor lattice QCD,''
  Phys.\ Rev.\ D {\bf 92}, no. 1, 014024 (2015)
  doi:10.1103/PhysRevD.92.014024
  [arXiv:1503.07839 [hep-lat]].
  %%CITATION = doi:10.1103/PhysRevD.92.014024;%%

%B->pi FNAL fT, f0
\bibitem{Bailey:2015nbd} 
  J.~A.~Bailey {\it et al.} [Fermilab Lattice and MILC Collaborations],
  %``$B\to\pi\ell\ell$ form factors for new-physics searches from lattice QCD,''
  Phys.\ Rev.\ Lett.\  {\bf 115}, no. 15, 152002 (2015)
  doi:10.1103/PhysRevLett.115.152002
  [arXiv:1507.01618 [hep-ph]].
  %%CITATION = doi:10.1103/PhysRevLett.115.152002;%%

%Bs->K HPQCD
\bibitem{Bouchard:2014ypa} 
  C.~M.~Bouchard, G.~P.~Lepage, C.~Monahan, H.~Na and J.~Shigemitsu,
  %``$B_s \to K \ell \nu$ form factors from lattice QCD,''
  Phys.\ Rev.\ D {\bf 90}, 054506 (2014)
  doi:10.1103/PhysRevD.90.054506
  [arXiv:1406.2279 [hep-lat]].
  %%CITATION = doi:10.1103/PhysRevD.90.054506;%%



%B->K

\bibitem{Bouchard:2013pna} 
  C.~Bouchard {\it et al.} [HPQCD Collaboration],
  %``Rare decay $B \to K \ell^+ \ell^-$ form factors from lattice QCD,''
  Phys.\ Rev.\ D {\bf 88}, no. 5, 054509 (2013)
  Erratum: [Phys.\ Rev.\ D {\bf 88}, no. 7, 079901 (2013)]
  doi:10.1103/PhysRevD.88.079901, 10.1103/PhysRevD.88.054509
  [arXiv:1306.2384 [hep-lat]].
  %%CITATION = doi:10.1103/PhysRevD.88.079901, 10.1103/PhysRevD.88.054509;%%
 
\bibitem{Bouchard:2013mia} 
  C.~Bouchard {\it et al.} [HPQCD Collaboration],
  %``Standard Model Predictions for $B \to K \ell^+ \ell^-$ with Form Factors from Lattice QCD,''
  Phys.\ Rev.\ Lett.\  {\bf 111}, no. 16, 162002 (2013)
  Erratum: [Phys.\ Rev.\ Lett.\  {\bf 112}, no. 14, 149902 (2014)]
  doi:10.1103/PhysRevLett.112.149902, 10.1103/PhysRevLett.111.162002
  [arXiv:1306.0434 [hep-ph]].
  %%CITATION = doi:10.1103/PhysRevLett.112.149902, 10.1103/PhysRevLett.111.162002;%%

\bibitem{Bailey:2015dka} 
  J.~A.~Bailey {\it et al.},
  %``$B\to Kl^+l^-$ decay form factors from three-flavor lattice QCD,''
  Phys.\ Rev.\ D {\bf 93}, no. 2, 025026 (2016)
  doi:10.1103/PhysRevD.93.025026
  [arXiv:1509.06235 [hep-lat]].
  %%CITATION = doi:10.1103/PhysRevD.93.025026;%%



%B->K* and Bs->phi
\bibitem{Horgan:2013hoa} 
  R.~R.~Horgan, Z.~Liu, S.~Meinel and M.~Wingate,
  %``Lattice QCD calculation of form factors describing the rare decays $B \to K^* \ell^+ \ell^-$ and $B_s \to \phi \ell^+ \ell^-$,''
  Phys.\ Rev.\ D {\bf 89}, no. 9, 094501 (2014)
  doi:10.1103/PhysRevD.89.094501
  [arXiv:1310.3722 [hep-lat]].
  %%CITATION = doi:10.1103/PhysRevD.89.094501;%%

\bibitem{Detmold:2015aaa} 
  W.~Detmold, C.~Lehner and S.~Meinel,
  %``$\Lambda_b \to p \ell^- \bar{\nu}_\ell$ and $\Lambda_b \to \Lambda_c \ell^- \bar{\nu}_\ell$ form factors from lattice QCD with relativistic heavy quarks,''
  Phys.\ Rev.\ D {\bf 92}, no. 3, 034503 (2015)
  doi:10.1103/PhysRevD.92.034503
  [arXiv:1503.01421 [hep-lat]].
  %%CITATION = doi:10.1103/PhysRevD.92.034503;%%


\bibitem{Straub:2015ica} 
  A.~Bharucha, D.~M.~Straub and R.~Zwicky,
  %``$B\to V\ell^+\ell^-$ in the Standard Model from light-cone sum rules,''
  JHEP {\bf 1608}, 098 (2016)
  doi:10.1007/JHEP08(2016)098
  [arXiv:1503.05534 [hep-ph]].
  %%CITATION = doi:10.1007/JHEP08(2016)098;%%

 
 
 
 


  \bibitem{Huber:2015sra} 
  T.~Huber, T.~Hurth and E.~Lunghi,
  %``Inclusive $ \overline{B}\to {X}_s{\ell}^{+}{\ell}^{-} $ : complete angular analysis and a thorough study of collinear photons,''
  JHEP {\bf 1506}, 176 (2015)
  doi:10.1007/JHEP06(2015)176
  [arXiv:1503.04849 [hep-ph]].
  %%CITATION = doi:10.1007/JHEP06(2015)176;%%

\bibitem{Ali:1996bm} 
  A.~Ali, G.~Hiller, L.~T.~Handoko and T.~Morozumi,
  %``Power corrections in the decay rate and distributions in $B \to X_s l^+ l^-$ in the Standard Model,''
  Phys.\ Rev.\ D {\bf 55}, 4105 (1997)
  doi:10.1103/PhysRevD.55.4105
  [hep-ph/9609449].
  %%CITATION = doi:10.1103/PhysRevD.55.4105;%%

\bibitem{Lee:2005pwa} 
  K.~S.~M.~Lee, Z.~Ligeti, I.~W.~Stewart and F.~J.~Tackmann,
  %``Universality and $m_X$ cut effects in $B \to  X_s l^+ l^-$,''
  Phys.\ Rev.\ D {\bf 74}, 011501 (2006)
  doi:10.1103/PhysRevD.74.011501
  [hep-ph/0512191].
  %%CITATION = doi:10.1103/PhysRevD.74.011501;%%

\bibitem{Benzke:2017}
M.~Benzke, M.~Fickinger, T.~Hurth and S.~Turczyk, to appear.
  
  
\bibitem{Hurth:2014vma} 
  T.~Hurth, F.~Mahmoudi and S.~Neshatpour,
  %``Global fits to $b \to s\ell\ell$ data and signs for lepton non-universality,''
  JHEP {\bf 1412}, 053 (2014)
  doi:10.1007/JHEP12(2014)053
  [arXiv:1410.4545 [hep-ph]].
  %%CITATION = doi:10.1007/JHEP12(2014)053;%%
  
  

 
   
%%%%%% LFV
\bibitem{Burdman:2001tf}
  G.~Burdman, E.~Golowich, J.~L.~Hewett and S.~Pakvasa,
  %``Rare charm decays in the standard model and beyond,''
  Phys.\ Rev.\ D {\bf 66} (2002) 014009
  doi:10.1103/PhysRevD.66.014009
  [hep-ph/0112235].
  %%CITATION = doi:10.1103/PhysRevD.66.014009;%%
  %219 citations counted in INSPIRE as of 09 Mar 2017


%%% Rare D decays
  
\bibitem{Golowich:2009ii}
  E.~Golowich, J.~Hewett, S.~Pakvasa and A.~A.~Petrov,
  %``Relating D0-anti-D0 Mixing and D0 ---> l+ l- with New Physics,''
  Phys.\ Rev.\ D {\bf 79} (2009) 114030
  doi:10.1103/PhysRevD.79.114030
  [arXiv:0903.2830 [hep-ph]].
  %%CITATION = doi:10.1103/PhysRevD.79.114030;%%
  %82 citations counted in INSPIRE as of 10 Mar 2017

\bibitem{Greub:1996wn}
  C.~Greub, T.~Hurth, M.~Misiak and D.~Wyler,
  %``The c ---> u gamma contribution to weak radiative charm decay,''
  Phys.\ Lett.\ B {\bf 382} (1996) 415
  doi:10.1016/0370-2693(96)00694-6
  [hep-ph/9603417].
  %%CITATION = doi:10.1016/0370-2693(96)00694-6;%%
  %87 citations counted in INSPIRE as of 10 Mar 2017

\bibitem{Fajfer:2001ad}
  S.~Fajfer, P.~Singer and J.~Zupan,
  %``The Rare decay D0 ---> gamma gamma,''
  Phys.\ Rev.\ D {\bf 64} (2001) 074008
  doi:10.1103/PhysRevD.64.074008
  [hep-ph/0104236].
  %%CITATION = doi:10.1103/PhysRevD.64.074008;%%
  %41 citations counted in INSPIRE as of 10 Mar 2017

\bibitem{Paul:2010pq}
  A.~Paul, I.~I.~Bigi and S.~Recksiegel,
  %``$D^0 -> \gamma \gamma$ and $D^0 -> \mu^+ \mu^-$ Rates on an Unlikely Impact of the Littlest Higgs Model with T-Parity,''
  Phys.\ Rev.\ D {\bf 82} (2010) 094006
   Erratum: [Phys.\ Rev.\ D {\bf 83} (2011) 019901]
  doi:10.1103/PhysRevD.83.019901, 10.1103/PhysRevD.82.094006
  [arXiv:1008.3141 [hep-ph]].
  %%CITATION = doi:10.1103/PhysRevD.83.019901, 10.1103/PhysRevD.82.094006;%%
  %26 citations counted in INSPIRE as of 10 Mar 2017

\bibitem{Nisar:2015gvd}
  N.~K.~Nisar {\it et al.} [Belle Collaboration],
  %``Search for the rare decay $D^0\to\gamma\gamma$ at Belle,''
  Phys.\ Rev.\ D {\bf 93} (2016) no.5,  051102
  doi:10.1103/PhysRevD.93.051102
  [arXiv:1512.02992 [hep-ex]].
  %%CITATION = doi:10.1103/PhysRevD.93.051102;%%
  %7 citations counted in INSPIRE as of 10 Mar 2017

\bibitem{Fajfer:2007dy}
  S.~Fajfer, N.~Kosnik and S.~Prelovsek,
  %``Updated constraints on new physics in rare charm decays,''
  Phys.\ Rev.\ D {\bf 76} (2007) 074010
  doi:10.1103/PhysRevD.76.074010
  [arXiv:0706.1133 [hep-ph]].
  %%CITATION = doi:10.1103/PhysRevD.76.074010;%%
  %44 citations counted in INSPIRE as of 10 Mar 2017
  
\bibitem{deBoer:2015boa}
  S.~de Boer and G.~Hiller,
  %``Flavor and new physics opportunities with rare charm decays into leptons,''
  Phys.\ Rev.\ D {\bf 93} (2016) no.7,  074001
  doi:10.1103/PhysRevD.93.074001
  [arXiv:1510.00311 [hep-ph]].
  %%CITATION = doi:10.1103/PhysRevD.93.074001;%%
  %17 citations counted in INSPIRE as of 10 Mar 2017  

\bibitem{Zhang:2010um}
  J.~M.~Zhang and G.~L.~Wang,
  %``Lepton-Number Violating Decays of Heavy Mesons,''
  Eur.\ Phys.\ J.\ C {\bf 71} (2011) 1715
  doi:10.1140/epjc/s10052-011-1715-1
  [arXiv:1003.5570 [hep-ph]].
  %%CITATION = doi:10.1140/epjc/s10052-011-1715-1;%%
  %32 citations counted in INSPIRE as of 10 Mar 2017
  
\bibitem{Li:2007kgb}
  H.~B.~Li and M.~Z.~Yang,
  %``Rare Semileptonic Decays of Heavy Mesons with Flavor SU(3) Symmetry,''
  Eur.\ Phys.\ J.\ C {\bf 59} (2009) 841
  doi:10.1140/epjc/s10052-008-0828-7
  [arXiv:0709.0979 [hep-ph]].
  %%CITATION = doi:10.1140/epjc/s10052-008-0828-7;%%
  %4 citations counted in INSPIRE as of 10 Mar 2017

\bibitem{Khodjamirian:2015dda}
  A.~Khodjamirian, T.~Mannel and A.~A.~Petrov,
  %``Direct probes of flavor-changing neutral currents in e$^{+}$ e$^{−}$-collisions,''
  JHEP {\bf 1511} (2015) 142
  doi:10.1007/JHEP11(2015)142
  [arXiv:1509.07123 [hep-ph]].
  %%CITATION = doi:10.1007/JHEP11(2015)142;%%
  %4 citations counted in INSPIRE as of 10 Mar 2017
  
\bibitem{Badin:2010uh}
  A.~Badin and A.~A.~Petrov,
  %``Searching for light Dark Matter in heavy meson decays,''
  Phys.\ Rev.\ D {\bf 82} (2010) 034005
  doi:10.1103/PhysRevD.82.034005
  [arXiv:1005.1277 [hep-ph]].
  %%CITATION = doi:10.1103/PhysRevD.82.034005;%%
  %26 citations counted in INSPIRE as of 10 Mar 2017

\bibitem{Lai:2016uvj}
  Y.-T.~Lai {\it et al.} [Belle Collaboration],
  %``Search for $D^{0}$ decays to invisible final states at Belle,''
  Phys.\ Rev.\ D {\bf 95} (2017) no.1,  011102
  doi:10.1103/PhysRevD.95.011102
  [arXiv:1611.09455 [hep-ex]].
  %%CITATION = doi:10.1103/PhysRevD.95.011102;%%
  %1 citations counted in INSPIRE as of 10 Mar 2017

  
 %%% Rare K decays

\bibitem{Buras:2015qea}
  A.~J.~Buras, D.~Buttazzo, J.~Girrbach-Noe and R.~Knegjens,
  %``$ {K}^{+}\to {\pi}^{+}\nu \overline{\nu} $ and $ {K}_L\to {\pi}^0\nu \overline{\nu} $ in the Standard Model: status and perspectives,''
  JHEP {\bf 1511} (2015) 033
  doi:10.1007/JHEP11(2015)033
  [arXiv:1503.02693 [hep-ph]].
  %%CITATION = doi:10.1007/JHEP11(2015)033;%%
  %76 citations counted in INSPIRE as of 15 Mar 2017
  
\bibitem{Buras:2015yca}
  A.~J.~Buras, D.~Buttazzo and R.~Knegjens,
  %``$ K\to \pi \nu \overline{\nu} $ and ε′/ε in simplified new physics models,''
  JHEP {\bf 1511} (2015) 166
  doi:10.1007/JHEP11(2015)166
  [arXiv:1507.08672 [hep-ph]].
  %%CITATION = doi:10.1007/JHEP11(2015)166;%%
  %30 citations counted in INSPIRE as of 15 Mar 2017
  
\bibitem{Artamonov:2009sz}
  A.~V.~Artamonov {\it et al.} [BNL-E949 Collaboration],
  %``Study of the decay $K^+\to\pi^+\nu \bar\nu$ in the momentum region $140 < P_\pi < 199$ MeV/c,''
  Phys.\ Rev.\ D {\bf 79} (2009) 092004
  doi:10.1103/PhysRevD.79.092004
  [arXiv:0903.0030 [hep-ex]].
  %%CITATION = doi:10.1103/PhysRevD.79.092004;%%
  %176 citations counted in INSPIRE as of 15 Mar 2017

\bibitem{Grossman:1997sk}
  Y.~Grossman and Y.~Nir,
  %``K(L) ---> pi0 neutrino anti-neutrino beyond the standard model,''
  Phys.\ Lett.\ B {\bf 398} (1997) 163
  doi:10.1016/S0370-2693(97)00210-4
  [hep-ph/9701313].
  %%CITATION = doi:10.1016/S0370-2693(97)00210-4;%%
  %264 citations counted in INSPIRE as of 16 Mar 2017

\bibitem{Smith:2014mla}
  C.~Smith,
  %``Rare K decays: Challenges and Perspectives,''
  arXiv:1409.6162 [hep-ph].
  %%CITATION = ARXIV:1409.6162;%%
  %8 citations counted in INSPIRE as of 17 Mar 2017

\bibitem{Isidori:2003ts}
  G.~Isidori and R.~Unterdorfer,
  %``On the short distance constraints from K(L,S) ---> mu+ mu-,''
  JHEP {\bf 0401} (2004) 009
  doi:10.1088/1126-6708/2004/01/009
  [hep-ph/0311084].
  %%CITATION = doi:10.1088/1126-6708/2004/01/009;%%
  %95 citations counted in INSPIRE as of 17 Mar 2017

\bibitem{LHCb:2016meg}
  The LHCb Collaboration [LHCb Collaboration],
  %``Updated limit for the decay $K_{\rm\scriptscriptstyle S}^0\rightarrow\mu^+\mu^-$,''
  LHCb-CONF-2016-012, CERN-LHCb-CONF-2016-012.
  %%CITATION = LHCB-CONF-2016-012, CERN-LHCB-CONF-2016-012;%%
  
\bibitem{Nayak:2016zed}
  A.~C.~Nayak and P.~Jain,
  %``Pion decay within the framework of Very Special Relativity,''
  arXiv:1610.01826 [hep-ph].
  %%CITATION = ARXIV:1610.01826;%%



  
  
\bibitem{TheMEG:2016wtm}
  A.~M.~Baldini {\it et al.} [MEG Collaboration],
  %``Search for the lepton flavour violating decay $\mu ^+ \rightarrow \mathrm {e}^+ \gamma $ with the full dataset of the MEG experiment,''
  Eur.\ Phys.\ J.\ C {\bf 76} (2016) no.8,  434
  doi:10.1140/epjc/s10052-016-4271-x
  [arXiv:1605.05081 [hep-ex]].
  %%CITATION = doi:10.1140/epjc/s10052-016-4271-x;%%
  %69 citations counted in INSPIRE as of 13 Mar 2017
\bibitem{Amhis:2016xyh}
  Y.~Amhis {\it et al.},
  %``Averages of $b$-hadron, $c$-hadron, and $\tau$-lepton properties as of summer 2016,''
  arXiv:1612.07233 [hep-ex].
  %%CITATION = ARXIV:1612.07233;%%
  %22 citations counted in INSPIRE as of 13 Mar 2017

\end{thebibliography}
